\documentclass[12pt]{article}

\usepackage{amsmath,amsfonts,amsthm,amssymb}
\usepackage{graphicx}
\usepackage[margin=1in]{geometry}
\usepackage[ruled,vlined]{algorithm2e}
\usepackage{subfigure}
\usepackage{wrapfig}
\usepackage{dsfont}
\usepackage{array}
\usepackage{nicematrix}
\newcolumntype{P}[1]{>{\centering\arraybackslash}p{#1}}
\newcolumntype{M}[1]{>{\centering\arraybackslash}m{#1}}

\usepackage{url}
\usepackage[breaklinks]{hyperref}

\newcommand{\expect}[1]{{\mathbb E}\left[{\displaystyle#1}\right]}

\usepackage[
    backend=biber,
    style=numeric,
    sorting=none,
    maxbibnames=99,
    minalphanames=3
]{biblatex}
\DeclareSourcemap{
  \maps[datatype=bibtex]{
    \map[overwrite]{
      \step[fieldsource=doi, final]
      \step[fieldset=url, null]
      \step[fieldset=eprint, null]
    }  
  }
}
\title{\Large{Deep Reinforcement Learning-based 
Rebalancing Policies \\
for Profit Maximization of Relay Nodes\\
in Payment Channel Networks\footnote{A version of this work appeared in the  \textit{4th International Conference on Mathematical Research for the Blockchain Economy (MARBLE 2023)} and received the best paper award.

This work was supported by a grant from JP Morgan Chase.
The authors would like to thank Leonidas Georgiadis, Nicholas Nordlund and Konstantinos Poularakis for helpful discussions.
}
\vspace{1cm}
}
}

\author{
	\textbf{Nikolaos Papadis} and \textbf{Leandros Tassiulas}\\
	Department of Electrical Engineering \& Institute for Network Science\\
	Yale University\\
	\texttt{\{nikolaos.papadis, leandros.tassiulas\}@yale.edu}
}
\date{}

\begin{document}

\maketitle

\begin{abstract}
Payment channel networks (PCNs) are a layer-2 blockchain scalability solution, with its main entity, the payment channel, enabling transactions between pairs of nodes ``off-chain,'' thus reducing the burden on the layer-1 network.
Nodes with multiple channels can serve as relays for multihop payments by providing their liquidity and withholding part of the payment amount as a fee.
Relay nodes might after a while end up with one or more unbalanced channels, and thus need to trigger a rebalancing operation.
In this paper, we study how a relay node can maximize its profits from fees by using the rebalancing method of submarine swaps.
We introduce a stochastic model to capture the dynamics of a relay node observing random transaction arrivals and performing occasional rebalancing operations, and express the system evolution as a Markov Decision Process.
We formulate the problem of the maximization of the node's fortune over time over all rebalancing policies, and approximate the optimal solution by designing a Deep Reinforcement Learning (DRL)-based rebalancing policy.
We build a discrete event simulator of the system and use it to demonstrate the DRL policy's superior performance under most conditions by conducting a comparative study of different policies and parameterizations.
Our work is the first to introduce DRL for liquidity management in the complex world of PCNs.
\end{abstract}

\newpage
\section{Introduction}
\label{sec:introduction}

Blockchain technology enables trusted interactions between untrusted parties, with financial applications like Bitcoin and beyond, but with also known scalability issues \cite{Cromanetal2016}. 
Payment channels are a layer-2 development towards avoiding the long confirmation times and high costs of the layer-1 network: they enable nodes that want to transact quickly, cheaply and privately to do so by depositing some balances to open a payment channel between themselves, and then trustlessly shifting the same total balance between the two sides without broadcasting their transactions and burdening the network.
Connected channels create a Payment Channel Network (PCN), via which two nodes not sharing a channel can still pay each other via a sequence of existing channels.
Intermediate nodes in the PCN function as relays: they forward the payment along its path and collect relay fees in return.
As transactions flow through the PCN, some channels get depleted, causing incoming transactions to fail because of insufficient liquidity on their path.
Thus, the need for channel rebalancing arises.

In this paper, we study the rebalancing mechanism of submarine swaps, which allows a blockchain node to exchange funds from on- to off-chain and vice versa.
Since a swap involves an on-chain transaction, it takes some time to complete.
Taking this into account, we formulate the following optimal rebalancing problem as a Markov Decision Process (MDP):
For a node relaying traffic across multiple channels, determine an optimal rebalancing strategy over time (i.e. when and how much to rebalance as a function of the transaction arrival rates observed from an unknown distribution and the confirmation time of an on-chain transaction), so that the node can keep its channels liquid and its profit from relay fees can be maximized.

More specifically, our \textit{contributions} are the following:
\begin{itemize}

    \item We develop a stochastic model that captures the dynamics of a relay node with two payment channels under two timescales: a continuous one for random discrete transaction arrivals in both directions from distributions unknown to the node, and a discrete one for dispatching rebalancing operations.

    \item We express the system evolution in our model as an MDP with continuous state and action spaces and time-dependent constraints on the actions, and formulate the problem of relay node profit maximization.

    \item We approximate the optimal policy of the MDP using Deep Reinforcement Learning (DRL) by appropriately engineering the states, actions and rewards and tuning a version of the Soft Actor-Critic algorithm.
    
    \item We develop a discrete event simulator of the system, and use it to evaluate the performance of the learning-based as well as other heuristic rebalancing policies under various transaction arrival conditions and demonstrate the superiority of our policy in a range of regimes.
    
\end{itemize}

In summary, our paper is the first to formally study the submarine swap rebalancing mechanism and to introduce a DRL-based method for channel rebalancing in particular, and for PCN liquidity management in general.

The remainder of the paper is organized as follows. 
In Sec. \ref{sec:background} we introduce the operation of payment channels and relay nodes, explain the need for rebalancing, and introduce the submarine swap rebalancing mechanism. 
In Sec. \ref{sec:problem-formulation} we describe our stochastic model of a relay node with two payment channels and write the profit maximization problem using an MDP.
In Sec. \ref{sec:heuristic-and-DRL-policies} we present heuristic policies as well as design a DRL-based algorithm for an approximately optimal solution to the problem, and in Sec. \ref{sec:evaluation} we describe the experimental setup and results. 
In Sec. \ref{sec:discussion} and \ref{sec:related-work} we discuss future directions and some related work.
Finally, Sec. \ref{sec:conclusion} concludes the paper.

\section{Background}
\label{sec:background}

\subsection{Payment channel operation}
\label{sec:payment-channel-operation}

\begin{wrapfigure}{R}{0.5\textwidth}
    \centering
    \includegraphics[width=0.48\textwidth]{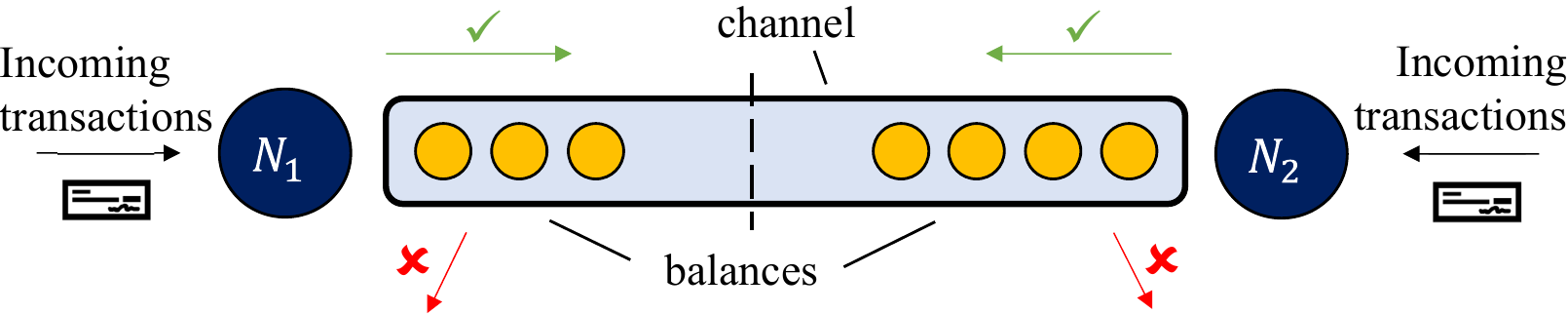}
    
    \caption{A payment channel between nodes $N_1$ and $N_2$ and current balances of 3 and 4}
    \label{fig:single-channel}
\end{wrapfigure}

A payment channel (Fig. \ref{fig:single-channel}) is created between two nodes $N_1$ and $N_2$ after they deposit some capital to a channel-opening on-chain transaction.
After this transaction is confirmed, the nodes can transact completely \textit{off-chain} (i.e. in the channel) without broadcasting their interactions to the layer-1 network, and without the risk of losing funds, thanks to a cryptographic safety mechanism.
The sum of their two balances in the channel remains constant and is called the channel capacity.
A transaction of amount $\alpha$ from $N_1$ to $N_2$ will succeed if the balance of $N_1$ at that moment suffices to cover it.
In this case, the balance of $N_1$ is reduced by $\alpha$ and the balance of $N_2$ is increased by $\alpha$.

\subsection{The role of relay nodes}
\label{sec:the-role-of-relay-nodes}

\begin{wrapfigure}{R}{0.5\textwidth}
    \centering
    \subfigure{
        \includegraphics[width=0.22\textwidth]{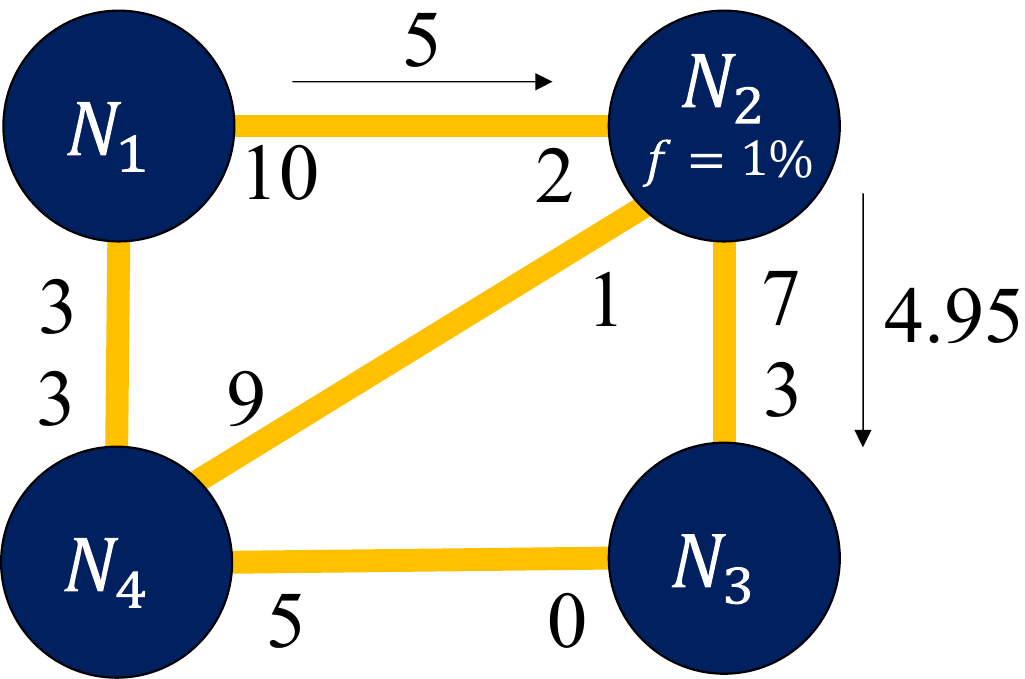}
        \label{fig:PCN_and_tx_with_fees_a}
    }
    \subfigure{
        \includegraphics[width=0.20\textwidth]{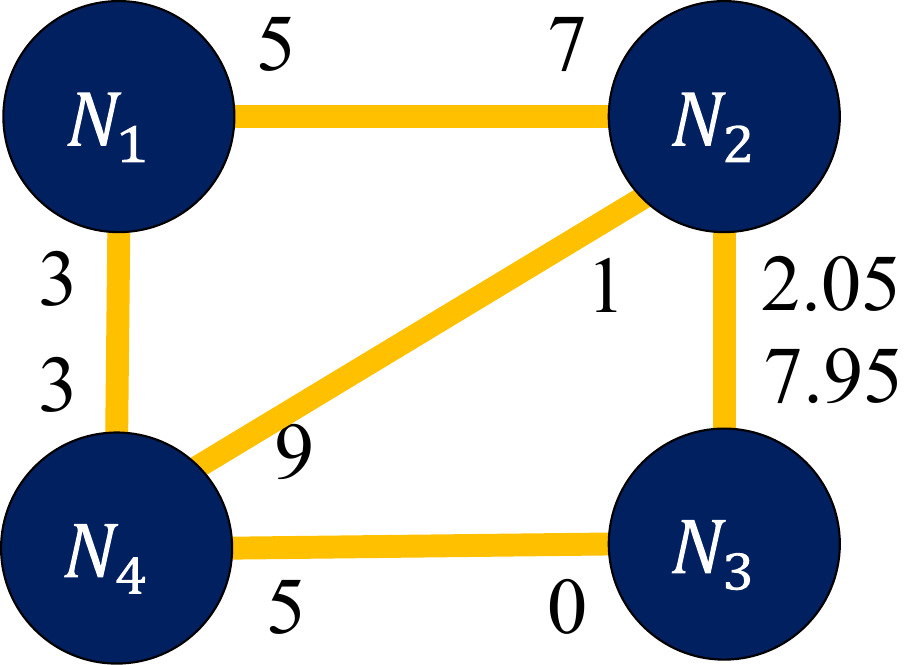}
        \label{fig:PCN_and_tx_with_fees_b}
    }
    \caption{Processing of a transaction in a payment channel network: before (left) and after (right)}
    \label{fig:PCN_and_tx_with_fees_before_and_after}
\end{wrapfigure}

As pairs of nodes create channels, a payment channel network (Fig. \ref{fig:PCN_and_tx_with_fees_before_and_after}) is formed, over which multihop payments are possible:
Consider a transaction of amount 5 from $N_1$ to $N_3$ via $N_2$.
Note that the amount 5 includes the fees that will have be paid on the way, e.g. 1\% at each intermediate node.
In the $N_1 N_2$ channel, $N_1$'s local balance is reduced by 5 and $N_2$'s local balance is increased by 5.
In the $N_2 N_3$ channel, $N_2$'s local balance is reduced by $5 - \text{fees} = 4.95$ and $N_3$'s local balance is increased by 4.95.
$N_2$'s total capital in all its channels before the transaction was $2+1+7=10$, while after it is $7+1+2.05=10.05$, so $N_2$ made a profit of 0.05 by acting as a relay.
If one of the outgoing balances did not suffice, then the transaction would fail end-to-end, thanks to a smart contract mechanism, the Hashed Time-Lock Contract (HTLC).
The role of relay nodes is fundamental for the continuous operation of a PCN.
The most prominent PCN currently is the Lightning Network \cite{Poon2016} built on top of Bitcoin.
More details on PCN operation can be found in \cite{Gudgeon2020a, Papadis2020}.

\subsection{The need for rebalancing}
\label{sec:the-need-for-rebalancing}

Depending on the demand a payment channel is facing in its two directions, funds might accumulate on one side and deplete on the other.
This might happen due to asymmetric demand inside single channels, the random nature of arrivals causing temporary depletions at specific times (e.g. when a large transaction arrives), or even symmetric demand between two endpoints of a multihop path which can cause imbalance due to fees withheld by intermediate nodes
(an example of this latter more subtle case is given in Appendix \ref{app:symmetric-depletion}).
The resulting imbalance is undesirable, as it leads to transaction failures and loss of profit from relay fees.
In fact, an entire PCN will stop being operational in finite time without external intervention, creating the need for rebalancing mechanisms.

\subsection{The submarine swap rebalancing mechanism}
\label{sec:submarine-swap-mechanism}

\begin{wrapfigure}{L}{0.5\textwidth}
    \centering
    \includegraphics[width=0.45\textwidth]{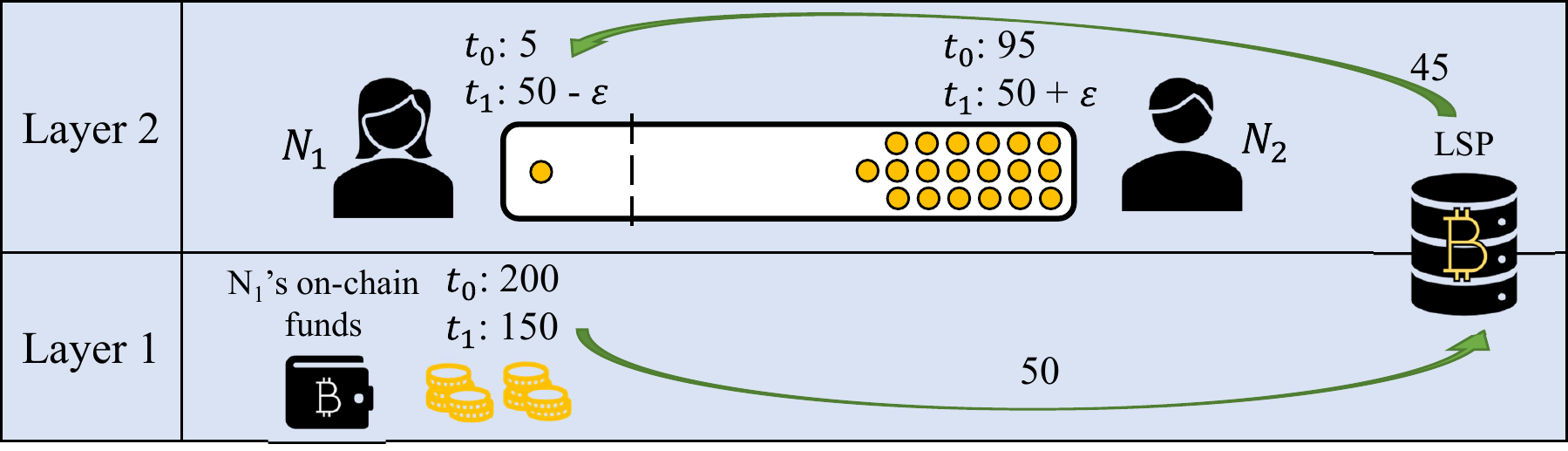}
    \caption{A submarine swap (swap-in)}
    \label{fig:submarine-swap-sketch}
\end{wrapfigure}

In this work, we study submarine swaps, introduced in \cite{Bosworth2018} and used commercially by Boltz\footnote{\url{https://boltz.exchange}} and Loop\footnote{\url{https://lightning.engineering/loop}}.
At a high level, a \textit{submarine swap} works as follows (Fig. \ref{fig:submarine-swap-sketch}): 
Node $N_1$ owns some funds in its channel with node $N_2$, and some funds on-chain.
At time $t_0$, the channel $N_1 N_2$ is almost depleted on $N_1$'s side (balance = 5). 
$N_1$ can start a \textit{swap-in} by paying an amount (50) to a Liquidity Service Provider (LSP) -- a wealthy node with access to both layers -- via an \textit{on-chain} transaction, and the LSP will give this amount back (reduced by a 10\% swap fee, so 45) to $N_1$ \textit{off-chain} via a path that goes through $N_2$.
The final amount that is added at $N_1$ (and subtracted at $N_2$) is $45-\varepsilon$ due to the relay fees spent on its way from the LSP.
Thus, at time $t_1$ the channel will be almost perfectly balanced.
The reverse process is also possible (a \textit{reverse submarine swap} or \textit{swap-out}) in order for a node to offload funds from its channel, by paying the server off-chain and receiving funds on-chain\footnote{In both cases, the layer-2 channel balances are altered with the help of an on-chain (layer-1, one layer below the channel) transaction, hence the characterization ``submarine.''}.

\begin{wrapfigure}{R}{0.5\textwidth}
    \centering
    \includegraphics[width=0.5\textwidth]{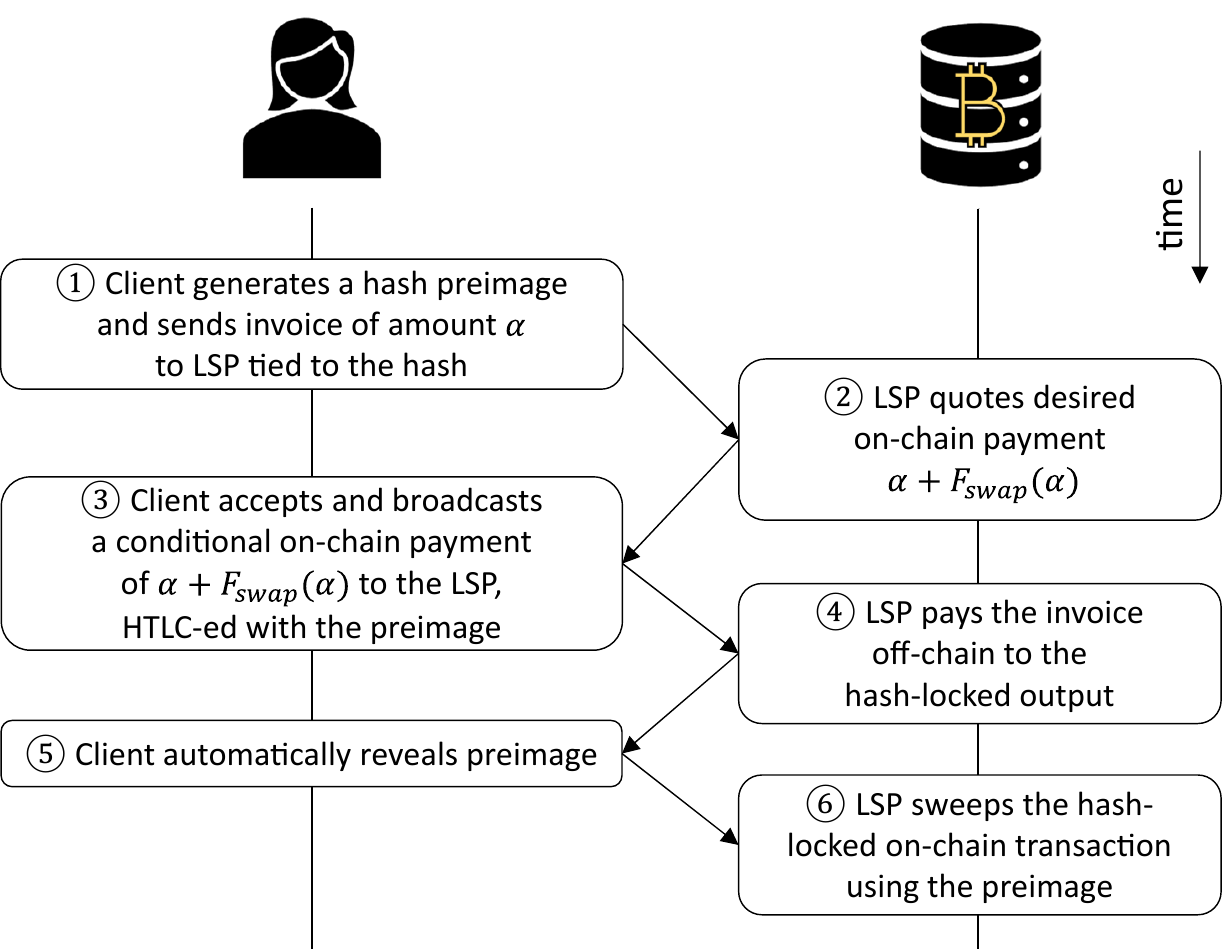}
    \caption{A swap-in step-by-step}
    \label{fig:submarine_swap_steps}
\end{wrapfigure}

A sketch of the technical protocol followed during a successful swap-in (a swap-out is similar) is shown in Fig. \ref{fig:submarine_swap_steps} and subsequently modeled in Sec. \ref{sec:a-swap-step-by-step}.
First, a node-client initiates the swap by generating a hash preimage, creating an invoice of the desired swap amount $r$ tied to this hash and with a certain expiration time $T_{\mathrm{exp}}$, and sends it to an LSP that is willing to make the exchange. 
The LSP then quotes what it wants to be paid on-chain in exchange for paying the client's invoice off-chain, say $\alpha + F_{\mathrm{swap}}(\alpha)$, where $F_{\mathrm{swap}}(\alpha)$ is the LSP's swap service fee.
If the client accepts the exchange rate, it creates a conditional on-chain payment of amount $\alpha + F_{\mathrm{swap}}(\alpha)$ to the LSP based on an HTLC with the same preimage as before
and broadcasts the payment to the blockchain network.
The payment can only be redeemed if the LSP knows the preimage, and the client will only reveal the preimage once it has received the LSP's funds on-chain.
Thus, the LSP pays the off-chain invoice.
This forces the client to reveal the preimage, and now the LSP can redeem the on-chain funds and the swap is complete.
The entire process happens trustlessly thanks to the HTLC mechanics.
More technical details can be found in \cite{submarine-swap-ion, loop-out-in-depth}.

There is an important tradeoff the node has to make, which is to strike a balance in terms of how often and how much it should rebalance: it can choose to not rebalance a lot to avoid paying swap fees, but then it forfeits profits from relay fees of transactions dropped due to imbalance.
On the other hand, it can choose to rebalance a lot so as not to drop any transaction, but then incurs high rebalancing fee costs.
This observation motivates us to study the problem of demand-aware and timely dispatching of swaps of the right amount by a node aiming to maximize its total fortune, which is presented in the next section.

\section{Problem formulation}
\label{sec:problem-formulation}

\subsection{System evolution}
\label{sec:system-evolution}

\begin{wrapfigure}{L}{0.36\textwidth}
    \centering
    \includegraphics[width=0.36\textwidth]{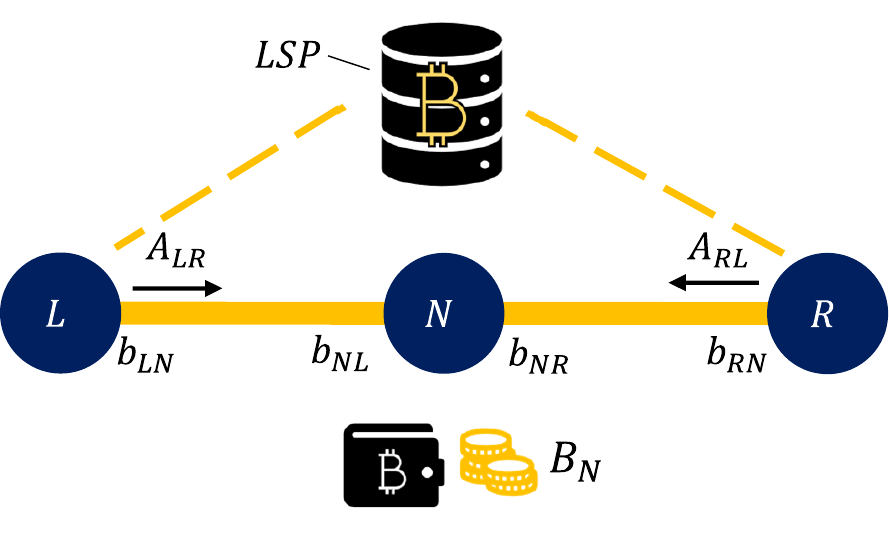}
    \caption{System model}
    \label{fig:system-model}
\end{wrapfigure}

In this section, we introduce a stochastic model of a PCN relay node $N$ that has two channels, one with node $L$ and one with node $R$, and wishes to maximize its profits from relaying payments from $L$ to $R$ and vice versa (Fig. \ref{fig:system-model}).
Let $b_{LN}(\tau), b_{NL}(\tau), b_{NR}(\tau), b_{RN}(\tau)$ be the channel balances and $B_N(\tau)$ be the on-chain amount of $N$ at time $\tau$.
Let $C_n$ be the total capacity of the channel $Nn$, $n \in \mathcal{N} \triangleq \{L, R \}$.
Events happen at two timescales: a continuous one for arriving transactions, and a discrete one for times when the node is allowed to rebalance.

\subsubsection{The transaction timescale}
\label{sec:the-transaction-timescale}

Transactions arrive as a marked point process and are characterized by their direction ($L$-to-$R$ or $R$-to-$L$), time of arrival and amount.
We consider node $N$ to not be the source or destination of any transactions itself, but rather to only act as a relay.
At each moment in continuous time (denoted by $\tau)$, (at most) one transaction arrives in the system.
All transactions are admitted, but some fail due to insufficient balances.

Let $f(\alpha)$ be the fee that a transaction of amount $\alpha$ pays to a node that relays it.
We assume all nodes charge the same fees. 
$f$ can be any fixed function with $f(0) = 0$. 
In practice, for $\alpha > 0$, $f(\alpha) = f_{\mathrm{base}} + f_{\mathrm{prop}} \cdot \alpha$, where the base fee $f_{\mathrm{base}}$ and the proportional fee $f_{\mathrm{prop}}$ are constants.
We state our model for a general relay fee function; however, as currently about 50\% of Lightning nodes have set the base fee to zero \cite{zero-base-fee}, in our experiments we too consider $f(\alpha) = f_{\mathrm{prop}} \cdot \alpha$.

Let $A_{LR}(\tau)$, $A_{RL}(\tau)$ be the externally arriving amounts coming from node $L$ in the $L$-to-$R$ direction and from node $R$ in the $R$-to-$L$ direction at time $\tau$ respectively, each drawn from a distribution that is fixed but unknown to node $N$.
An arriving transaction of amount $A_{LR}(\tau) = \alpha$ is feasible if and only if there is enough balance in the $L$-to-$R$ direction in both channels, i.e. $b_{LN}(\tau) \geq \alpha$ and $b_{NR} \geq \alpha - f(\alpha)$, and similarly for the $R$-to-$L$ direction.
The successfully admitted and processed amounts by node $N$ at time $\tau$ are\footnote{Since in the sequel we focus on the discrete and sparse time scale of the periodic times at which the node rebalances, we make the fair assumption (as e.g. in \cite{Bai2022}) that off-chain transactions are processed instantaneously across their entire path and do not fail in their subsequent steps after they cross the two channels (if a transaction were to fail outside the two channels, it can be viewed as of zero value by the system).}:
\begin{align}
    S_{LR}(\tau) &= 
        \begin{cases} 
            A_{LR}(\tau)    & \text{, if } A_{LR}(\tau) \leq b_{LN}(\tau)  \text{ and } A_{LR}(\tau) - f(A_{LR}(\tau)) \leq b_{NR}(\tau)\\
            0                   & \text{, otherwise} \\
        \end{cases}
    \\
    S_{RL}(\tau) &= 
        \begin{cases} 
            A_{RL}(\tau)    & \text{, if } A_{RL}(\tau) \leq b_{RN}(\tau)  \text{ and } A_{RL}(\tau) - f(A_{RL}(\tau)) \leq b_{NL}(\tau)\\
            0                   & \text{, otherwise} \\
        \end{cases}
\end{align}

Then the profit of node $N$ at time $\tau$ is $f(S_{LR}(\tau)) + f(S_{RL}(\tau))$, and the lost fees (from transactions that potentially failed to process) are 
$f\bigl( A_{LR}(\tau) - S_{LR}(\tau) \bigr) + f\bigl( A_{RL}(\tau) - S_{RL}(\tau) \bigr)$.

\noindent
The balance processes at time $\tau$ evolve as follows:
\begin{align}
    b_{LN}(\tau) &\rightarrow b_{LN}(\tau) + (S_{RL}(\tau) - f(S_{RL}(\tau))) - S_{LR}(\tau) \\
    b_{NL}(\tau) &\rightarrow b_{NL}(\tau) + S_{LR}(\tau) - (S_{RL}(\tau) - f(S_{RL}(\tau))) \\
    b_{NR}(\tau) &\rightarrow b_{NR}(\tau) + S_{RL}(\tau) - (S_{LR}(\tau) - f(S_{LR}(\tau))) \\
    b_{RN}(\tau) &\rightarrow b_{RN}(\tau) + (S_{LR}(\tau) - f(S_{LR}(\tau))) - S_{RL}(\tau)
\end{align}
The on-chain amount $B_N(\tau)$ is not affected by the processing of off-chain transactions.

\subsubsection{The rebalancing decision (control) timescale}
\label{sec:the-rebalancing-timescale}

The evolution of the system can be controlled by node $N$ using submarine swap rebalancing operations. 
Rebalancing may start at times $t_i = i \cdot T_{\mathrm{check}}$, $i=0,1,...$, and takes a (fixed) time $T_{\mathrm{conf}}$ to complete (on average 10 minutes for Bitcoin)\footnote{In practice, completion happens when the miners solve the random puzzle and produce the Proof-of-Work for the next block that includes the rebalancing transaction. 
The time for this to happen fluctuates, though only slightly, so we use a fixed value for the sake of tractability.}.
We consider the case where $T_{\mathrm{check}} \geq T_{\mathrm{conf}}$ (to avoid having concurrent rebalancing operations in the same channel that could be combined into one).

The system state is defined only for the discrete rebalancing decision timescale as the collection of the off- and on-chain balances:
\begin{equation}
    S(t_i) = \bigl( b_{LN}(t_i), b_{NL}(t_i), b_{NR}(t_i), b_{RN}(t_i), B_N(t_i) \bigr)
    \label{eqn:state-definition}
\end{equation}

At each time $t_i$, node $N$ can decide to request a swap-in or a swap-out in each channel. 
Call the respective amounts $r^{\mathrm{in}}_L(t_i), r^{\mathrm{out}}_L(t_i), r^{\mathrm{in}}_R(t_i), r^{\mathrm{out}}_R(t_i)$.
At any time $t_i$, in a given channel, either a swap-in or a swap-out or nothing will requested, but not both a swap-in and a swap-out\footnote{The other two nodes ($L$ and $R$) are considered passive, i.e. they perform no swap operations themselves.}.

Let $F_{\mathrm{swap}}^{\mathrm{in}}(\alpha)$ and $F_{\mathrm{swap}}^{\mathrm{out}}(\alpha)$ be the swap fees that the LSP charges for an amount $\alpha$ for a swap-in and a swap-out respectively, where $F_{\mathrm{swap}}^{\mathrm{in}}(\cdot)$ and $F_{\mathrm{swap}}^{\mathrm{out}}(\cdot)$ are any nonnegative functions with $F_{\mathrm{swap}}(0) = 0$.
For ease of exposition, we let all types of fees the node will have to pay (relay fees for the off-chain part, on-chain miner fees, server fees) be part of the above swap fees, and be the same for both swap-ins and swap-outs when a net amount $r_{\mathrm{net}}$ is transferred from on- to off-chain or vice versa: $F_{\mathrm{swap}}^{\mathrm{in}}(r_{\mathrm{net}}) = F_{\mathrm{swap}}^{\mathrm{out}}(r_{\mathrm{net}}) = F_{\mathrm{swap}}(r_{\mathrm{net}}) \triangleq r_{\mathrm{net}} F + M$, where the proportional part $F$ includes the server fee and off-chain relay fees, and $M$ includes the miner fee and potential base fees.

Note that the semantics of the swap amounts $r$ are such that they represent the amount that will move \textit{in the channel} (and not necessarily the net change in the node's fortune). 
As a result of this convention and based on the swap operation as described in section \ref{sec:a-swap-step-by-step}, the amount $r^{\mathrm{in}}$ of a swap-in coincides with the net amount $r^{\mathrm{in}}_{\mathrm{net}}$ by which the node's fortune decreases (as $r^{\mathrm{in}}$ does not include the swap fee), while the amount $r^{\mathrm{out}}$ of a swap-out includes the swap fee and the net amount by which the node's fortune decreases is $r^{\mathrm{out}}_{\mathrm{net}} = \phi^{-1}(r^{\mathrm{out}})$, where $\phi(r_{\mathrm{net}}) \triangleq r_{\mathrm{net}} + F_{\mathrm{swap}}(r_{\mathrm{net}})$, and $\phi^{-1}$ is the generalized inverse function of $\phi(\cdot)$ (it always exists: $\phi^{-1}(y) = \min\{x \in \mathbb{N}: \phi(x) = y\}$).
For our $F_{\mathrm{swap}}(\cdot)$ it is $\phi(r_{\mathrm{net}}) = r_{\mathrm{net}} (1+F) + M$ for $r_{\mathrm{net}} > 0$, $\phi(0)= 0$, so $\phi^{-1}(y) = (y-M) / (1+F)$ for $y>0$ and $\phi^{-1}(0) = 0$.

\subsubsection{A submarine swap step-by-step}
\label{sec:a-swap-step-by-step}

We now describe how a rebalancing operation on the $Nn$ channel is affecting the system state.
First, we describe a swap-in of amount $r_n^{\mathrm{in}}$ initiated by node $N$ to refill $N$'s local balance in the $Nn$ channel:

\begin{itemize}
    \item At time $t_i$, node $N$ locks the net rebalancing amount plus fees and subtracts it from its on-chain funds: $B_N \rightarrow B_N - (r_n^{\mathrm{in}} + F_{\mathrm{swap}}^{\mathrm{in}}(r_n^{\mathrm{in}}))$

    \item At time $t_i + T_{\mathrm{conf}}$, the on-chain transaction is confirmed, so the LSP sends a payment of $r_n^{\mathrm{in}}$ to node $N$ off-chain\footnote{The LSP is a well-connected node owning large amounts of liquidity, so we reasonably assume that it can always find a route from itself to $N$, possibly via splitting the amount across multiple paths.}.
    The rebalancing payment reaches node $n$:
    
    If $b_{nN} \leq r_n^{\mathrm{in}}$ (i.e. $n$ does not have enough balance to forward it), then the off-chain payment fails. The on-chain funds are unlocked\footnote{In practice, the on-chain funds are unlocked after a time $T_{\mathrm{exp}}$ to prevent malicious clients from requesting many swaps from an LSP and then defaulting. 
    However, since we are concerned with online and cooperative clients with on-chain amounts usually quite larger than the amounts in their channels (and thus than their swaps), and also there is currently a community effort to reduce or even eliminate $T_{\mathrm{exp}}$, we ignore it.} and refunded back to the on-chain amount: $B_N \rightarrow B_N + (r_n^{\mathrm{in}} + F_{\mathrm{swap}}^{\mathrm{in}}(r_n^{\mathrm{in}}))$
    
    Otherwise (if the transaction is feasible), $n$ forwards the payment to $N$: $b_{nN} \rightarrow b_{nN} - r_n^{\mathrm{in}}$ and $b_{Nn} = b_{Nn} + r_n^{\mathrm{in}}$
\end{itemize}

A swap-out of amount $r_n^{\mathrm{out}}$, initiated by node $N$ to offload some of its local balance to the chain, works as follows:

\begin{itemize}
    \item At time $t_i$, node $N$ locks the net rebalancing amount plus fees and sends it to the LSP via the off-chain network: $b_{Nn} \rightarrow b_{Nn} - r_n^{\mathrm{out}}$. 
    Note that $r_n^{\mathrm{out}}$ includes the fees.
    
    \item At time $t_i + T_{\mathrm{conf}}$, the on-chain transaction is confirmed, so node $N$ receives the funds on-chain: $B_N \rightarrow B_N +  \phi^{-1}(r_n^{\mathrm{out}})$, and the funds are also unlocked in the channel and pushed towards the remote balance: $b_{nN} \rightarrow b_{nN} + r_n^{\mathrm{out}}$
\end{itemize}

\subsubsection{Rebalancing constraints}
\label{sec:rebalancing-constraints}

Based on the steps just described, swap operations will succeed if and only if their amounts satisfy the following constraints:

\begin{itemize}
\item
Rebalancing amounts must be non-negative:
\begin{equation}
r^{\mathrm{in}}_n(t_i), r^{\mathrm{out}}_n(t_i) \geq 0 \text{ for all } i \in \mathbb{N}, n \in \mathcal{N}
\label{constraint:nonnegative}
\end{equation}

\item
A swap-in and a swap-out cannot be requested in the same channel at the same time\footnote{This fact allows us to express the decision per channel as a single variable taking both positive and negative values, instead of two non-negative variables. We do so in Section \ref{sec:DRL-algorithm-design}, but we retain two action variables per channel in the present section for the sake of clarity.}:
\begin{align}
    r^{\mathrm{in}}_n(t_i) \cdot r^{\mathrm{out}}_n(t_i) = 0 \text{ for all } i \in \mathbb{N}, n \in \mathcal{N} \label{constraint:not-both-swaps-in-channel}
\end{align}

\item
The swap-out amounts (which already include the swap fees) must be greater than the fees themselves:
\begin{align}
    r^{\mathrm{out}}_n(t_i) - F_{\mathrm{swap}}^{\mathrm{out}}(r^{\mathrm{out}}_n(t_i)) &\geq 0 \text{ for all } i \in \mathbb{N}, n \in \mathcal{N} \label{constraint:min-swap-out}
\end{align}

\item
The respective channel balances must suffice to cover the swap-out amounts (which already include the swap fees):
\begin{align}
    r^{\mathrm{out}}_n(t_i) \leq b_{Nn}(t_i) \text{ for all } i \in \mathbb{N}, n \in \mathcal{N} \label{constraint:swap-out-enough-remote}
\end{align}

\item
The on-chain balance must suffice to cover the total swap-in amount plus fees: 
\begin{equation}
    \sum_{n \in \mathcal{N}} \bigl( r^{\mathrm{in}}_n(t_i) + F_{\mathrm{swap}}^{\mathrm{in}}(r^{\mathrm{in}}_n(t_i)) \bigr) \leq B_N(t_i)   \text{ for all } i \in \mathbb{N}
\label{constraint:coupled-swap-in}
\end{equation}
\end{itemize}

\subsubsection{State evolution equations}
\label{sec:state-evolution-equations}

Now we are able to write the complete state evolution equations. 
The amounts added to each balance due to successful transactions during the interval $(t_i, t_{i+1})$ are
\begin{align}
    d_{NL}^{(t_i, t_{i+1})} \triangleq 
    \int_{\tau \in (t_i, t_{i+1})} \bigg( 
    S_{LR}(\tau) - \left( S_{RL}(\tau) - f(S_{RL}(\tau)) \right) 
    \bigg)  d\tau \\
    d_{NR}^{(t_i, t_{i+1})} \triangleq 
    \int_{\tau \in (t_i, t_{i+1})} \bigg( 
    S_{RL}(\tau) - \left( S_{LR}(\tau) - f(S_{LR}(\tau)) \right) 
    \bigg) d\tau 
\end{align}
and $d_{nN}^{(t_i, t_{i+1})} \triangleq - d_{Nn}^{(t_i, t_{i+1})}$.
Then for actions taken subject to the constraints 
\eqref{constraint:nonnegative}--\eqref{constraint:coupled-swap-in}, the state evolves as follows:
\begin{align}
    b_{nN}(t_{i+1}) &= b_{nN}(t_i) 
        + d_{nN}^{(t_i, t_{i+1})}
        - (r_n^{\mathrm{in}}(t_i) - z_n(t_i))
        + r_n^{\mathrm{out}}(t_i)
        \\
    b_{Nn}(t_{i+1}) &= b_{Nn}(t_i) 
        + d_{Nn}^{(t_i, t_{i+1})}
        + (r_n^{\mathrm{in}}(t_i) - z_n(t_i))
        - r_n^{\mathrm{out}}(t_i)
        \\
    B_N(t_{i+1}) &= B_N(t_i)
        - \sum_{n \in \mathcal{N}} \bigg( 
         r_n^{\mathrm{in}}(t_i) - F^{\mathrm{in}}_{\mathrm{swap}}(r_n^{\mathrm{in}}(t_i)) \bigg)
        + \sum_{n \in \mathcal{N}} \phi^{-1}(r_n^{\mathrm{out}}(t_i))
        + \sum_{n \in \mathcal{N}}w_n(t_i)  
\end{align}
where $z_n(t_i)$ and $w_n(t_i)$ are the refunds of the swap-in amount off- and on-chain respectively in case a swap-in operation fails:
\begin{align}
    z_n(t_i) &= r_n^{\mathrm{in}}(t_i) \mathds{1} \{ b_{nN}(t_i) + d_{nN}^{(t_i, t_i + T_{\mathrm{conf}})} < r_n^{\mathrm{in}}(t_i) \} \label{eqn:refund-z} \\
    w_n(t_i) &= z_n(t_i) + F_{\mathrm{swap}}^{\mathrm{in}}(z_n(t_i)) \label{eqn:refund-w}
\end{align}

\subsection{Writing the problem as a Markov Decision Process}
\label{sec:writing-as-an-MDP}

The objective function the node wishes to maximize in the real world is its \textit{total fortune both in the channels and on-chain}.
The fortune increase due to the action (the 4-tuple) $r(t_i)$ taken at step $t_i$ is:
\begin{equation}
    D(t_i, r(t_i)) \triangleq \bigg( \sum_{n \in \mathcal{N}} b_{Nn}(t_{i+1}) + B_{N}(t_{i+1}) \bigg)
    - \bigg( \sum_{n \in \mathcal{N}} b_{Nn}(t_{i}) + B_{N}(t_{i}) \bigg)
    \label{eqn:objective-fortune-increase}
\end{equation}

Equivalently, the node can minimize the total fee cost, which comes from two sources: from lost fees because of dropped transactions\footnote{Note that we assume the node knows not only about the transactions that reach it, but also about the transactions that are supposed to reach it but never do because of insufficient remote balances.
This is not strictly true in practice, but the node can approximate it by observing the transactions during an interval in which the remote balances are both big enough so that no incoming transaction would fail, and create an estimate based on this observation.
}, and from fees paid for rebalancing operations:
\begin{align}
\begin{split}
L(t_i, r(t_i))      = &\int_{\tau \in (t_i, t_{i+1})} \bigl( f(A_{LR}(\tau) - S_{LR}(\tau)) + f(A_{RL}(\tau) 
            - S_{RL}(\tau)) \bigr) d\tau \\
            &+ \sum_{n \in \mathcal{N}} \bigl( F_{\mathrm{swap}}^{\mathrm{in}}(r^{\mathrm{in}}_n(t_i)) + F_{\mathrm{swap}}^{\mathrm{out}}(r^{\mathrm{out}}_n(t_i)) \bigr)
\label{eqn:objective-fee-loss}
\end{split}
\end{align}

The two objectives at each timestep sum to 
$\int_{\tau \in (t_i, t_{i+1})} \left( f(A_{LR}(\tau) + f(A_{RL}(\tau) \right) d\tau$
(the fees that would be collected by the node if the total arriving amount had been processed), which is a quantity independent of the control action, and therefore maximizing the total fortune and minimizing the total fee cost are equivalent.

A control policy $\pi = \{(t_i, r^\pi(t_i))\}_{i \in \mathbb{N}}$ consists of the times $t_i$ and the corresponding actions $r^\pi(t_i) = \left( r^{\mathrm{in}}_L(t_i), r^{\mathrm{out}}_L(t_i), r^{\mathrm{in}}_R(t_i), r^{\mathrm{out}}_R(t_i) \right)$ taken from the set of allowed actions $\mathcal{R} = [0, C_L]^2 \times [0, C_R]^2$, and belongs to the set of admissible policies
\begin{equation*}
    \Pi = \bigl\{ \{(t_i, r(t_i))\}_{i \in \mathbb{N}} \text{ such that } r(t_i) \in \mathcal{R} \text{ for all } i \in \mathbb{N} \bigr\}
    \label{eqn:admissible-policies}
\end{equation*}

Ultimately, the goal of node $N$ is to find a rebalancing policy that maximizes the long-term average expected fortune increase $D$ (equivalently, minimizes the long-term average expected fee cost $L$) over all admissible rebalancing policies:
\begin{align}
&\underset{\pi \in \Pi}{\text{maximize}} \lim_{H \rightarrow \infty} \frac{1}{t_H} \sum_{i=0}^H \expect{D(t_i, r^\pi(t_i))} 
\end{align}
subject to the constraints
\eqref{constraint:nonnegative}--\eqref{constraint:coupled-swap-in}.

\section{Heuristic and deep reinforcement learning-based rebalancing policies}
\label{sec:heuristic-and-DRL-policies}

In this section, we describe the steps we took in order to apply DRL to approximately solve the formulated MDP.
We first outline two heuristic policies, which we will use later to benchmark our DRL-based solution.

\subsection{Heuristic policies}
\label{sec:heuristic-policies}

\begin{algorithm}[ht]
\LinesNumbered
\SetKwInOut{Parameter}{Parameters}
\SetKwFor{Every}{every}{do}{}
\KwIn{$state$ as in Eq. \eqref{eqn:state-definition}}
\Parameter{$T_{\mathrm{check}}$, \textit{low, high}}

\Every{$T_{\mathrm{check}}$} {
    \ForEach{neighbor $n \in \mathcal{N}$} {
        $midpoint = C_n \cdot (low + high) /2 $
        
        \If{$b_{Nn} < low \cdot C_n$}{
            Swap-in amount = $midpoint - b_{Nn}$
        }    
        \ElseIf{$b_{Nn} > high \cdot C_n$}{
            Swap-out amount = $b_{Nn} - midpoint$
        }
        \Else{
            Perform no action
        }
    }
}
\caption{Autoloop rebalancing policy}
\label{alg:autoloop}
\end{algorithm}

Autoloop \cite{autoloop-1, autoloop-2} is a policy that allows a node to schedule automatic swap-ins (resp. swap-outs) if its local balance falls below a minimum (resp. rises above a maximum) threshold expressed as a percentage of the channel's capacity (Alg. \ref{alg:autoloop})\footnote{The original Autoloop algorithm defines the thresholds in terms of inbound liquidity in a node's channel. 
We adopt an equivalent balance-centric view instead.}.
The initiated swap is of amount equal to the difference of the local balance from the midpoint, i.e. the average of the two thresholds.
We expect Autoloop to be suboptimal with respect to profit maximization in certain cases, as it does not take the expected demand into account and thus possibly performs rebalancing at times when it is not necessary.

This motivates us to define another heuristic policy that incorporates the empirical demand information.
We call this policy Loopmax (Alg. \ref{alg:loopmax}), as its goal is to rebalance with the maximum possible amount and as infrequently as possible (without sacrificing transactions), based on the demand at each time.
Loopmax keeps track of the total arriving amounts, and estimates the net change of each balance per unit time using the difference of the total amounts that arrived in each direction:
\begin{align}
\begin{split}
    \hat{A}_{LN}^{\mathrm{net}}(\tau) = - \hat{A}_{NL}^{\mathrm{net}}(\tau) 
    \triangleq \dfrac{1}{\tau}
    \int_{t \in [0, \tau]} \bigg( A_{RL}(t) - f(A_{RL}(t)) - A_{LR}(t) \bigg) dt
    \label{estimate:net-demand-L}
\end{split}
\end{align}
\begin{align}
\begin{split}
    \hat{A}_{RN}^{\mathrm{net}}(\tau) = - \hat{A}_{NR}^{\mathrm{net}}(\tau) 
    \triangleq \dfrac{1}{\tau}
    \int_{t \in [0, \tau]} \bigg( A_{LR}(t) - f(A_{LR}(t)) - A_{RL}(t) \bigg) dt
    \label{estimate:net-demand-R}
\end{split}
\end{align}

For each channel, we first calculate the estimated time to depletion ($ETTD$) or saturation ($ETTS$) of the channel, depending on the direction of the net demand and the current balances, and using this time we dispatch a swap of the appropriate type not earlier than $T_{\mathrm{check}} + T_{\mathrm{conf}}$ before depletion/saturation, and of the maximum possible amount. 
The rationale is that if e.g. $ETTD \geq T_{\mathrm{check}} + T_{\mathrm{conf}}$, the policy can leverage this fact to postpone starting a swap until the next check time, since until then no transactions will have been dropped.
If $ETTD < T_{\mathrm{check}} + T_{\mathrm{conf}}$ though, the policy should act now, as otherwise it will end up dropping transactions during the following interval of  $T_{\mathrm{check}} + T_{\mathrm{conf}}$.
The maximum possible swap-out is constrained by the local balance at that time, while the maximum possible swap-in is constrained by the remote balance at that time\footnote{Actually, it is constrained by the remote balance at the time of the swap-in's completion. We will improve this later using estimates of future balances.} and the on-chain amount: an on-chain amount of $B_N$ can support (by including fees) a net swap-in amount of at most $\phi^{-1}(B_N)$.

Compared to Autoloop, Loopmax has the advantage that it rebalances only when it is absolutely necessary and can thus achieve savings in swap fees.
On the other hand, Loopmax's aggressiveness can lead it to extreme rebalancing decisions when traffic is quite skewed in a particular direction (e.g. it can do a swap-in of almost the full capacity, which is very likely to fail due to randomness in the transaction arrivals).
A small modification we can use on top of Alg. \ref{alg:loopmax} to alleviate this is to define certain safety margins of liquidity that Loopmax should always leave intact on each side of the channel, so that incoming transactions do not find it depleted due to a large pending swap.

\begin{algorithm}[ht]
\LinesNumbered
\SetKwInOut{Parameter}{Parameters}
\SetKwFor{On}{on}{do}{}
\SetKwFor{Every}{every}{do}{}
\KwIn{$state$ as in Eq. \eqref{eqn:state-definition}}
\Parameter{$T_{\mathrm{check}}$}

\Every{$T_{\mathrm{check}}$}{
    Update $\{\hat{A}_{Nn}^{\mathrm{net}}\}_{n \in \mathcal{N}}$ according to Eqs. \eqref{estimate:net-demand-L}--\eqref{estimate:net-demand-R}
    
    \ForEach{neighbor $n \in \mathcal{N}$} {
        \If{$\hat{A}_{Nn}^{\mathrm{net}} < 0$}{
            $ETTD = b_{Nn} / | \hat{A}_{Nn}^{\mathrm{net}} |$ /* estimated time to depletion */
            
            \If{$ETTD < T_{\mathrm{check}} + T_{\mathrm{conf}}$}{
                Swap-in amount = $\max \{\phi^{-1}(B_N), b_{nN} \}$ /* maximum possible swap in */
            }
            \Else{
            Perform no action
            }
        }    
        \ElseIf{$\hat{A}_{Nn}^{\mathrm{net}} > 0$}{
            $ETTS = b_{nN} / \hat{A}_{Nn}^{\mathrm{net}}$ /* estimated time to saturation */
            
            \If{$ETTS < T_{\mathrm{check}} + T_{\mathrm{conf}}$}{
                Swap-out amount = $b_{Nn}$ /* maximum possible swap out */
            }
            \Else{
            Perform no action
            }
        }
        \Else{
            Perform no action
        }
    }
}
\caption{Loopmax rebalancing policy}
\label{alg:loopmax}
\end{algorithm}

\subsection{Deep reinforcement learning algorithm design}
\label{sec:DRL-algorithm-design}

Having formulated the problem as an MDP, we now need to find an (approximately) optimal policy.
The problem is challenging for a number of reasons:
\begin{itemize}
    \item The problem dynamics are not linear.
    \item The state and action spaces are continuous and thus tabular approaches are not applicable.
    \item There are time-dependent constraints on the actions.
    \item Choosing to not rebalance at a specific time requires special treatment, as otherwise the zero action will be sampled from a continuous action space with zero probability.
\end{itemize}

To tackle these challenges, we resort to approximate methods, and specifically Reinforcement Learning (RL).
In the standard RL framework, an agent makes decisions based on a policy that is represented as a probability distribution over states and actions: $p: p(s,a) \rightarrow [0,1]$, with $p(s,a)$ being the probability that action $a$ will be taken when the environment is in state $s$.
Since our problem has continuous state and action spaces and the policy cannot be stored in tabular form, we need to use function approximation techniques.
Neural networks serve well the role of function approximators in many applications \cite{alizadeh2016}.
Some algorithms appropriate for this type of problems are Deep Deterministic Policy Gradient (DDPG) \cite{DDPG-paper} and Soft Actor-Critic (SAC) \cite{SAC-paper-2}.
We decided to use the latter, as DDPG is known to exhibit extreme brittleness and hyperparameter sensitivity \cite{Duan16}.

We now describe our methodology around how we engineer our DRL algorithm based on the vanilla SAC in order to arrive at a solution that deals with all the above challenges.

For the RL agent's environment, we use as state the five balances (off- and on-chain) and the estimates of the remote balances at the time of the swap completion, each normalized appropriately: by the respective channel's capacity, or by a total target fortune in the on-chain amount's case.
Thus, our state space is $[0,1]^7$.
As actions, instead of the 4-tuple of Section \ref{sec:problem-formulation}, we use a 2-tuple $(r_L, r_R)$, i.e. a single variable for each channel that can take both positive (swap-in) and negative (swap-out) values.
Raw actions are sampled from the entire continuous action space; 
before the raw action is applied, it undergoes some processing described in the sequel.

As mentioned, an action with a coordinate equal to zero would be selected with zero probability.
In reality, though, performing zero rebalancing in a channel when a swap is not necessary is important for maximizing the fortune/minimizing the costs, and an action the agent should learn to apply.
To this end, if the raw action coordinate is less than a threshold $\rho_0$ (e.g. 20\%) of the channel capacity, we force the respective applied action to be zero.
This way, we make the zero action selectable with positive probability, and at the same time prevent the agent from performing swaps too small in size (which would increase the cost).

Moreover, in order to guide the algorithm to respect the constraints, we perform an additional processing step.
The vanilla SAC algorithm \cite{SAC-paper-2} operates on an action space that is a compact subset of $\mathbb{R}^k$ for all decision times.
In our case, though, the allowed actions vary due to the time-dependent constraints \eqref{constraint:nonnegative}--\eqref{constraint:coupled-swap-in}.
We therefore define the action space to be $[-1,1]^2$, where each coordinate denotes the percentage not of the entire channel capacity, but of the maximum amount available for the respective type of swap at that moment.
We now focus on deriving these maximum amounts from the constraints.

All constraints are decoupled per channel, except for \eqref{constraint:coupled-swap-in}.
However, we observe that given some traffic, mostly in the $L$-to-$R$ direction or mostly in the $R$-to-$L$ direction or equal in both directions, the local balances of node $N$ will either deplete in one channel and accumulate in the other, or accumulate in both, but never both deplete.
Thus, a swap-in in both channels in general will not be a good action.
Therefore, for the RL solution's purposes we can split \eqref{constraint:coupled-swap-in} into two constraints, one for each channel, with the right-hand side of each being the entire amount $B_N(t_i)$.
In case the agent does take the not advisable decision of swap-ins in both channels and their sum exceeds the on-chain amount, one of the two will simply fail.

Another useful observation is that when a swap-in is about to complete time $T_{\mathrm{conf}}$ after it  commenced, the remote balance in the respective channel needs to suffice (otherwise the swap-in will fail and a refund will be triggered as in Eqs. \eqref{eqn:refund-z}--\eqref{eqn:refund-w}):
\begin{align}
    r^{\mathrm{in}}_n(t_i) &\leq b_{nN}(t_i) + d_{nN}^{(t_i, t_i + T_{\mathrm{conf}})} \text{ for all } i \in \mathbb{N}, n \in \mathcal{N} 
    \label{constraint:future-con-initial}
\end{align}
Although \eqref{constraint:future-con-initial} are not hard constraints when the decision is being made like the ones of Section \ref{sec:rebalancing-constraints}, we would like to guide the agent to respect them.
An obstacle is that the swap-in decision is made at time $t_i$, when the node does not yet know the arriving amount $d_{nN}^{(t_i, t_i + T_{\mathrm{conf}})}$.
To approximate the right-hand side of \eqref{constraint:future-con-initial} in terms of quantities known at time $t_i$, we can use the difference of the total (and not the successful as in $d_{nN}$'s definition) amounts that arrived in each direction (Eqs. \eqref{estimate:net-demand-L}--\eqref{estimate:net-demand-R}):
\begin{align}
\begin{split}
    b_{nN}(t_i) + d_{nN}^{(t_i, t_i + T_{\mathrm{conf}})} \approx
    \hat{b}_{nN}(t_i + T_{\mathrm{conf}}) \triangleq    \bigl( \min \{ b_{nN}(t_i) + \hat{A}_{nN}^{\mathrm{net}} \cdot T_{\mathrm{conf}}, C_n  \} \bigr)^+
    \label{estimate:future-balance-2}
\end{split}
\end{align}

A better estimate can be obtained by using the empirical amounts that succeeded in either direction:
\begin{equation}
    \hat{S}_{LR}(\tau)
    \triangleq \dfrac{1}{\tau}
    \int_{t \in [0, \tau]} S_{LR}(t) dt
    \qquad \text{and} \qquad
    \hat{S}_{RL}(\tau)
    \triangleq \dfrac{1}{\tau}
    \int_{t \in [0, \tau]} S_{RL}(t) dt
    \label{estimate:successful-amounts}
\end{equation}

Then the amount $\hat{S}_{LR}$ (resp. $\hat{S}_{RL}$) will be flowing in the $L$-to-$R$ (resp. $R$-to-$L$) direction for either the entire duration of $T_{\mathrm{conf}}$, or until one of the balances in the respective direction is depleted:
\begin{align}
\begin{split}
    \hat{b}_{LN}(t_i + T_{\mathrm{conf}}) \triangleq \bigg( \min \bigg\{ b_{LN}(t_i)
    &- \hat{S}_{LR}(t_i) \min \big\{ T_{\mathrm{conf}}, \dfrac{b_{LN}}{\hat{S}_{LR}(t_i)}, \dfrac{b_{NR}}{\hat{S}_{LR}(t_i)} \big\} \\
    &+ (1 - f_{\mathrm{prop}}) \hat{S}_{RL}(t_i) \min \big\{ T_{\mathrm{conf}}, \dfrac{b_{RN}}{\hat{S}_{RL}(t_i)}, \dfrac{b_{NL}}{\hat{S}_{RL}(t_i)} \big\}
    , C_L  \bigg\} \bigg)^+
    \label{estimate:future-balance-4-L}
\end{split}
\end{align}
\begin{align}
\begin{split}
    \hat{b}_{RN}(t_i + T_{\mathrm{conf}}) \triangleq \bigg( \min \bigg\{ b_{RN}(t_i)
    &- \hat{S}_{RL}(t_i) \min \big\{ T_{\mathrm{conf}}, \dfrac{b_{RN}}{\hat{S}_{RL}(t_i)}, \dfrac{b_{NL}}{\hat{S}_{RL}(t_i)} \big\} \\
    &+ (1 - f_{\mathrm{prop}}) \hat{S}_{LR}(t_i) \min \big\{ T_{\mathrm{conf}}, \dfrac{b_{LN}}{\hat{S}_{LR}(t_i)}, \dfrac{b_{NR}}{\hat{S}_{LR}(t_i)} \big\}
    , C_R  \bigg\} \bigg)^+
    \label{estimate:future-balance-4-R}
\end{split}
\end{align}

Thus, the approximate version of \eqref{constraint:future-con-initial} becomes:
\begin{equation}
    r^{\mathrm{in}}_n(t_i) \leq \hat{b}_{nN}(t_i + T_{\mathrm{conf}}) \text{ for all } i \in \mathbb{N}, n \in \mathcal{N} 
    \label{constraint:future-con-final}
\end{equation}

Note that we have given the agent more flexibility compared to Autoloop and Loopmax: it is allowed to perform swap-ins of amount bigger than the one allowed by the current balances, under the expectation that by the time of their completion the balances will be adequate.

Now we can write all constraints \eqref{constraint:nonnegative}--\eqref{constraint:coupled-swap-in}, \eqref{constraint:future-con-final} in terms of the 2-tuple $(r_L, r_R)$ as follows:
\begin{equation*}
    r_n \in \left[-b_{Nn}, -\rho_{\mathrm{min}}^{\mathrm{out}}\right] \cup \left[0, \min \{\hat{b}_{nN}(t_i + T_{\mathrm{conf}}), \phi^{-1}(B_N(t_i)), C_n\}\right], n \in \mathcal{N}
    \label{eqn:action-bounds-separate}
\end{equation*}
where $\rho_{\mathrm{min}}^{\mathrm{out}} \triangleq M / (1-F)$ is the minimum solution of \eqref{constraint:min-swap-out}.

If $\rho_0 C_n \gg \rho_{\mathrm{min}}^{\mathrm{out}}$, which should hold in practice as $\rho_{\mathrm{min}}^{\mathrm{out}}$ is very small, we can write
\begin{equation}
    r_n \in \left[-b_{Nn}, \min \{\hat{b}_{nN}(t_i + T_{\mathrm{conf}}), \phi^{-1}(B_N(t_i)), C_n\} \right], n \in \mathcal{N}
    \label{eqn:action-bounds-unified}
\end{equation}

The final mapping of raw actions (sampled from the distribution on the entire action space) to the finally applied actions is shown in Table \ref{table:action-processing}.

\begin{table}
\caption{Mapping of raw actions sampled from the learned distribution to final swap amounts requested for channel $Nn$, $n \in \mathcal{N}$}
\label{table:action-processing}
\centering
\makebox[\textwidth]{\begin{tabular}{ |c|c|c| } 
\hline
Raw action $r_n \in [-1,1]$     & Corresponding absolute amount $\Tilde{r}_n$    & Final requested swap amount \\    
\hline
$r_n < 0$   & $|r_n| b_{Nn}$  & swap out $\Tilde{r}_n \mathds{1}\{\Tilde{r}_n \geq \rho_0 C_n\}$  \\ 
$r_n \geq 0$   & $r_n \min \{\hat{b}_{nN}(t_i + T_{\mathrm{conf}}), \phi^{-1}(B_N(t_i)), C_n\}$  & swap in $\Tilde{r}_n \mathds{1}\{\Tilde{r}_n > \rho_0 C_n\}$  \\ 
\hline  
\end{tabular}}
\end{table}

We craft the reward signal to guide the agent towards optimizing the objective: 
we add the node's fortune increase \eqref{eqn:objective-fortune-increase} until the next check time, subtract the fee losses from transactions dropped until the next check time, and also subtract a fixed penalty for every swap the algorithm initiates and which eventually fails.
A high-level description of the most important components of the final learning process is given in Alg. \ref{alg:RebEL}.
We call the emerging policy ``RebEL'': Rebalancing Enabled by Learning.

\begin{algorithm}[ht]
\LinesNumbered
\SetKwInOut{Parameter}{Parameters}
\SetKwFor{On}{on}{do}{}
\SetKwFor{Every}{every}{do}{}
\KwIn{$state$ as in Eq. \eqref{eqn:state-definition}}
\Parameter{$T_{\mathrm{check}}$, various learning parameters, penalty}

\Every{$T_{\mathrm{check}}$}{
    Update estimates $\hat{S}_{LR}$, $\hat{S}_{RL}$ and $\hat{b}_{LN}$, $\hat{b}_{RN}$ according to Eqs. \eqref{estimate:successful-amounts}--\eqref{estimate:future-balance-4-R}
    
    Perform SAC gradient step to update policy distribution as in \cite{SAC-paper-2} based on replay memory
    
    Fetch $state \in [0,1]^7$
    
    Sample $rawAction$ from $[-1,1]^2$ according to policy distribution
    
    $processedAction = process(rawAction)$ where $process(\cdot)$ is described in Table \ref{table:action-processing}
    
    Apply $processedAction$ and wait for its completion
    
    $reward$ = fortuneAfter $-$ fortuneBefore $-$ lostFees $-$ penalty $\cdot$ \#OfFailedSwaps
    
    Fetch $nextState \in [0,1]^7$
    
    Store transition $(state, rawAction, reward$, $nextState)$ to replay memory
}
\caption{RL algorithm for RebEL policy}
\label{alg:RebEL}
\end{algorithm}

\vspace{-1em}
\section{Evaluation}
\label{sec:evaluation}

\subsection{Simulator}
\label{sec:simulator}

In order to evaluate the performance of different rebalancing policies, we build a discrete event simulator of a relay node with two payment channels and rebalancing capabilities using Python SimPy \cite{simpy}. 
The simulator treats each channel as a resource allowed to undergo at most one active swap at a time, and allows for parameterization of the initial balances, the transaction generation distributions (frequency, amount, number) in both directions, the different fees, the swap check and confirmation times, the rebalancing policy and its parameters\footnote{The code is publicly available at \url{https://github.com/npapadis/payment-channel-rebalancing}.}.

\subsection{Experimental setup}
\label{sec:experimental-setup}

We simulate a relay node with two payment channels, each of a capacity of \$1000 split equally between the channel's nodes.
Transactions arrive from both sides as Poisson processes.
We evaluate policies Autoloop, Loopmax and RebEL defined in Sec. \ref{sec:heuristic-and-DRL-policies}, as well as the \textit{None} policy that never performs any rebalancing.
We use $T_{\mathrm{check}} = T_{\mathrm{conf}} = 10$ minutes, miner fee $M = \$2$/on-chain transaction (tx), swap fee $F = 0.5\%$, 0.3 and 0.7 as the low and high liquidity thresholds of Autoloop, and 2 minutes worth of estimated traffic as safety margins for Loopmax.
We run all experiments on a regular consumer laptop.

We experimented with different hyperparameters for the original SAC algorithm\footnote{We used the PyTorch implementation in \url{https://github.com/pranz24/pytorch-soft-actor-critic}.} as well as for RebEL parameters and reward shapes, and settled with the ones shown in Appendix \ref{app:hyperparameters-and-rewards}.
We performed experiments for the transaction amount distribution being Uniform in $[0, 50]$ and Gaussian with mean 25 and standard deviation 20, and the results were very similar. 
Therefore, all plots shown below are for the Gaussian amounts.

\subsection{Results}
\label{sec:results}

\subsubsection{The role of fees}
\label{sec:experiments-wrt-fees}

Current median fee rates for transaction forwarding are in the order of $3 \cdot 10^{-5}$ (\$/\$) or 0.003\%\footnote{\url{https://1ml.com/statistics}}, while swap server fees are in the order of 0.5\%\footnote{\url{https://lightning.engineering/loop}} and miner fees are in the order of 2 \$/tx\footnote{\url{https://ycharts.com/indicators/bitcoin\_average\_transaction\_fee}}.
In order to see if a relay node can make a profit with such fees, we perform the following back-of-the-envelope calculation:
A swap-in of amount $r$ will cost the node $rF+M$ in fees and will enable traffic of at most value $r$ to be processed, which will yield profits $r f_{\mathrm{prop}}$ from relay fees.
Therefore, the swap-in cannot be profitable if $rF+M \geq r f_{\mathrm{prop}}$.
Solving this inequality, we see that no positive amount $r$ can be profitable if $f_{\mathrm{prop}} \leq F$, while if $f_{\mathrm{prop}} > F$ a necessary (but not sufficient) condition for profitability is $r > M / (f_{\mathrm{prop}}-F)$.
The respective inequality for a swap-out of amount $r$ is $r - \frac{r-M}{1+F} \geq r f_{\mathrm{prop}}$, which shows that for $f_{\mathrm{prop}} \leq \frac{F}{1+F}$ no amount can be profitable and for $f_{\mathrm{prop}} > \frac{F}{1+F}$ a necessary condition for profitability is that $r > \frac{M}{f_{\mathrm{prop}}(1+F)-F}$.
\textit{With the current fees, we are in the non-profitable regime.}
Although the above inequalities are short-sighted in that they focus only on a specific action time, they do confirm the observation made by both the Lightning and the academic communities \cite{Beres2020v1} that in order for relay nodes to be a profitable business, relay fees have to increase.

\begin{wrapfigure}{R}{0.65\textwidth}
    \centering
    \subfigure[Total fortune over time under high demand skewed in the $L$-to-$R$ direction]{
        \includegraphics[width=0.3\textwidth]{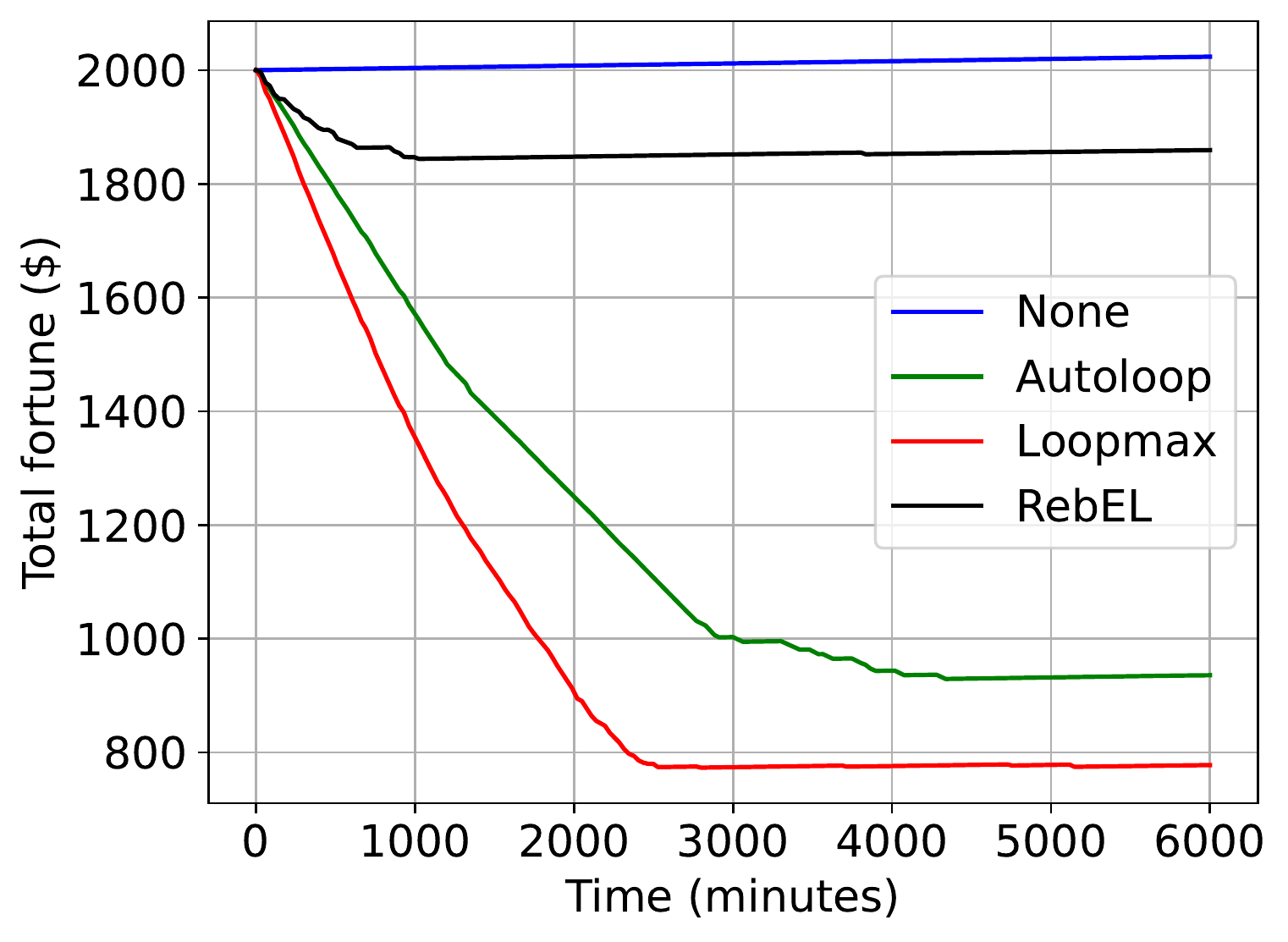}        \label{fig:results_202_total_fortune_over_time}
    }
    \subfigure[Total final fortune under high demand skewed in the $L$-to-$R$ direction for different values of the proportional relay fee $f_{\mathrm{prop}}$]{
        \includegraphics[width=0.3\textwidth]{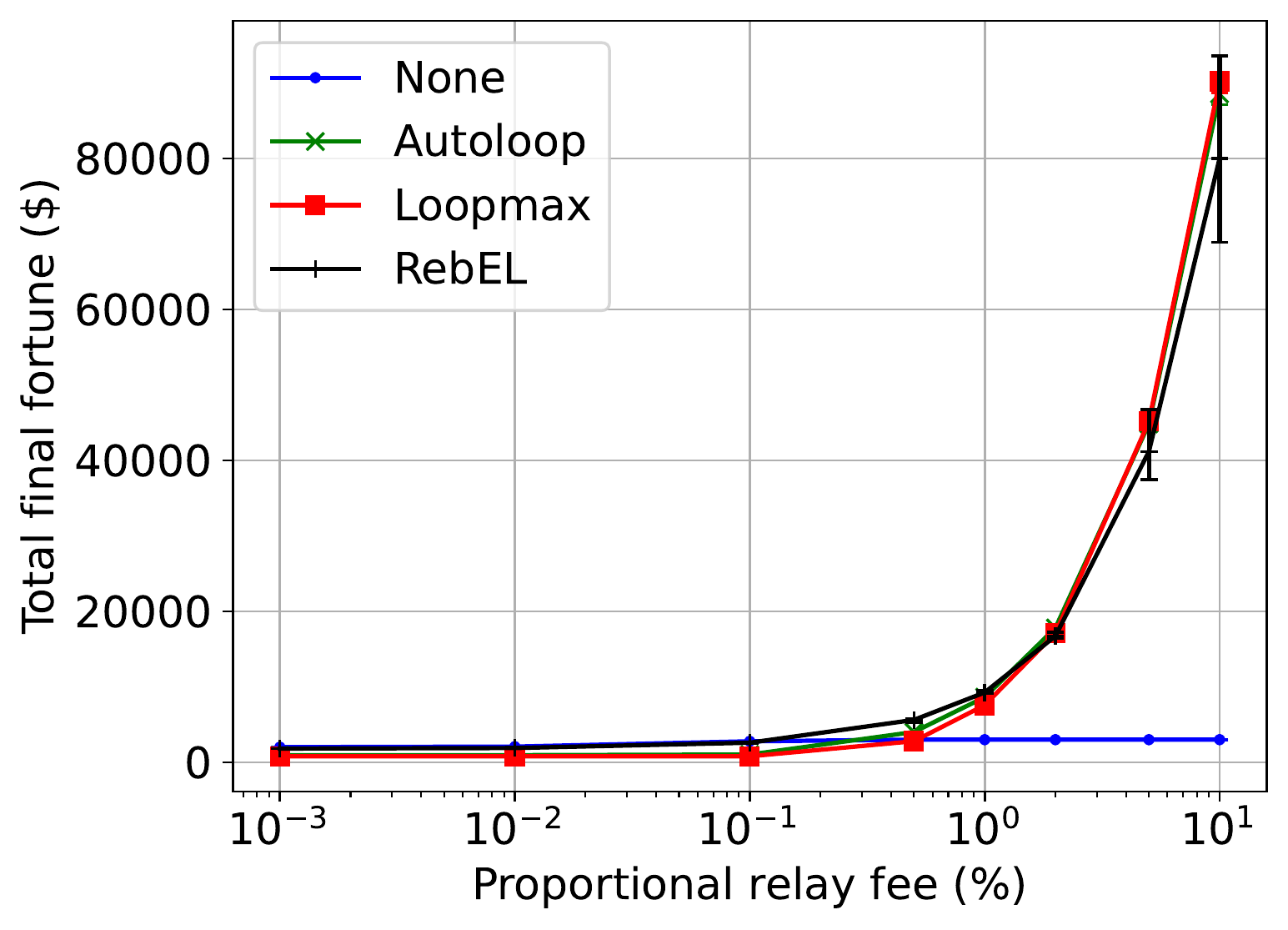}        \label{fig:results_212_final_fortune_wrt_proportional_fee}
    } \\
    \subfigure[Total fortune over time under low demand skewed in the $L$-to-$R$ direction]{
        \includegraphics[width=0.3\textwidth]{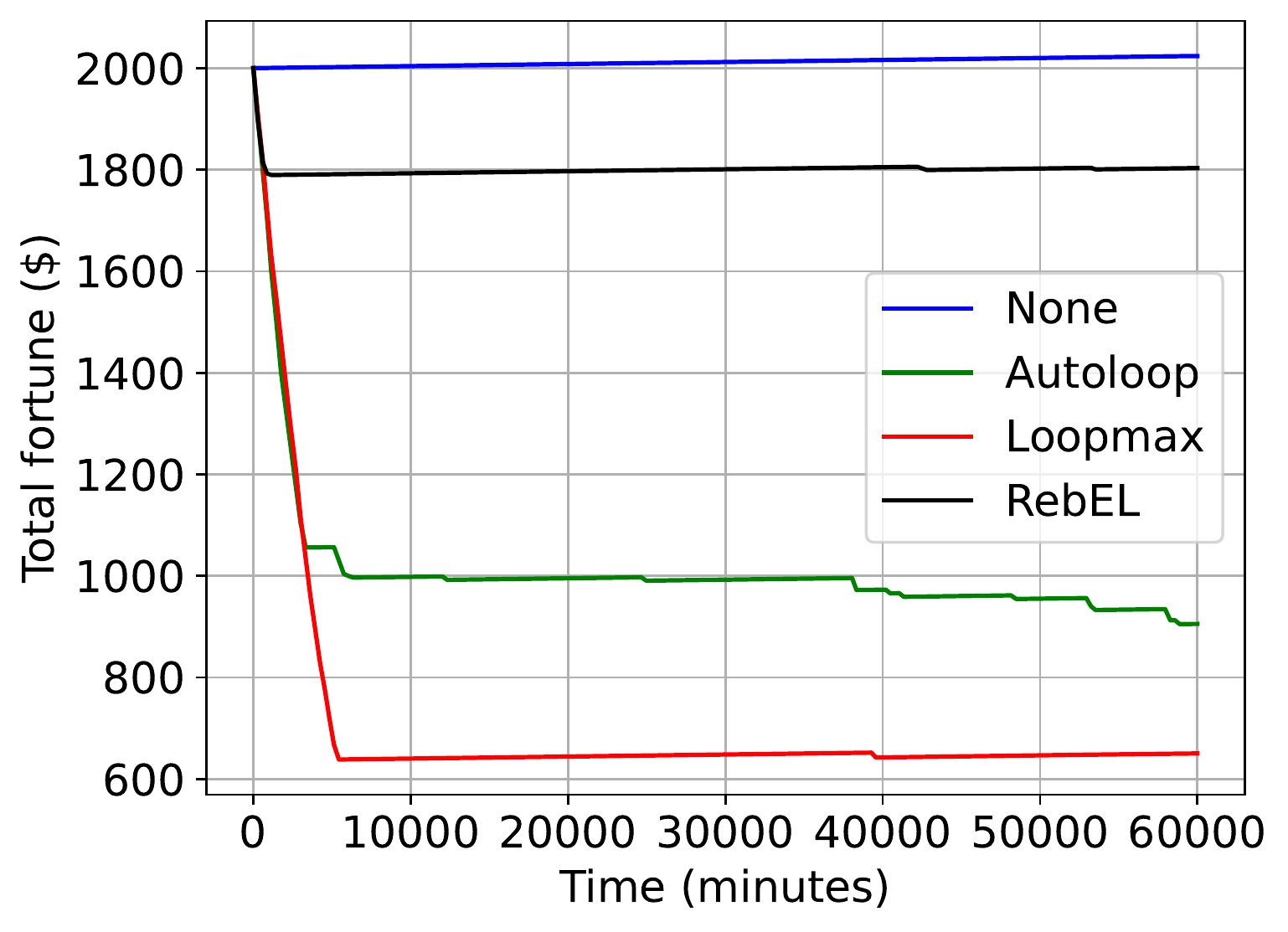}        \label{fig:results_204_total_fortune_over_time}
    }
    \subfigure[Total final fortune under low demand skewed in the $L$-to-$R$ direction for different values of the proportional relay fee $f_{\mathrm{prop}}$]{
        \includegraphics[width=0.3\textwidth]{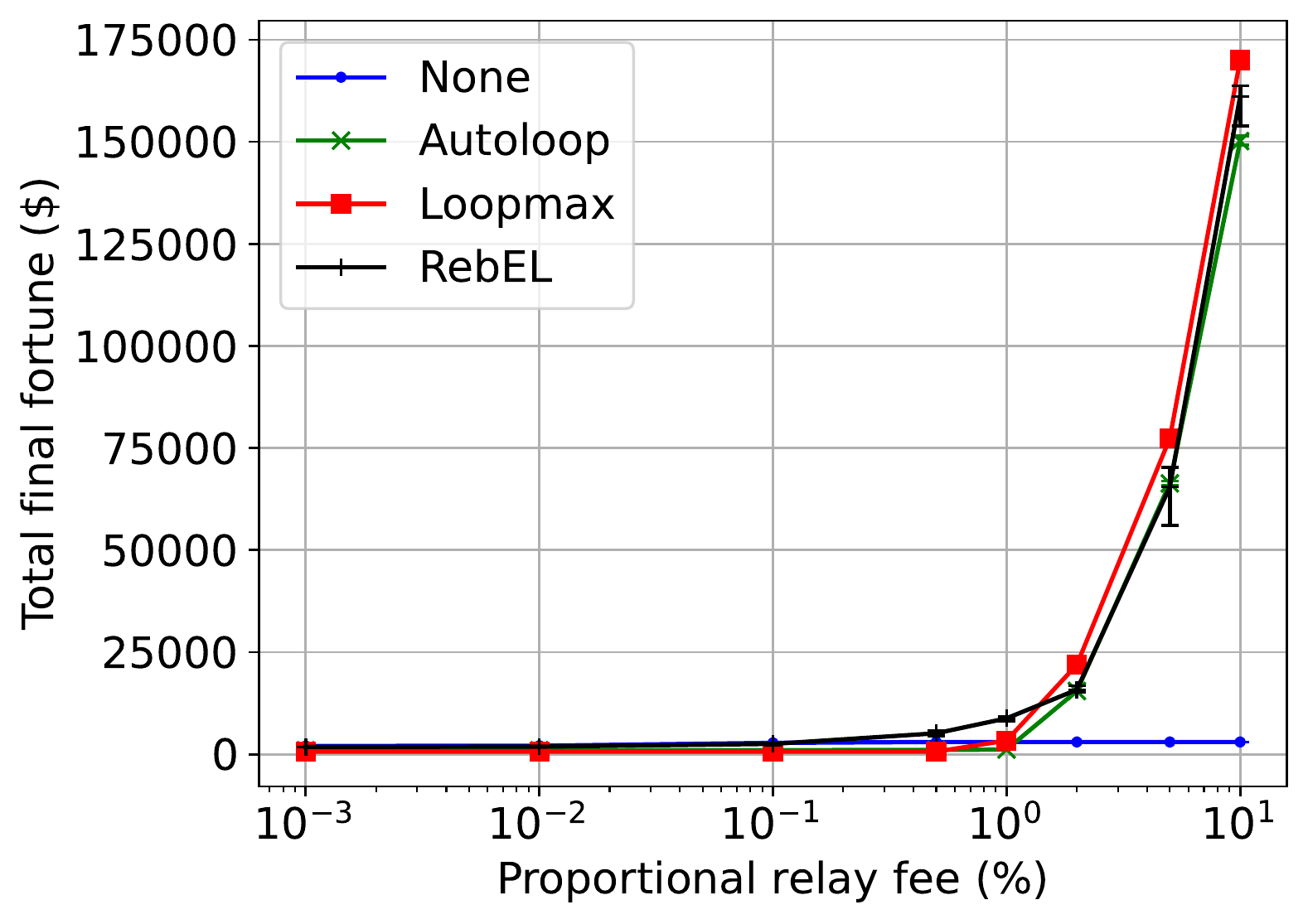}
        \label{fig:results_214_final_fortune_wrt_proportional_fee}
    }
    \caption{Experiments with different proportional relay fee $f_{\mathrm{prop}}$}
    \label{fig:final-fortune-varying-fees}
\end{wrapfigure}

We now perform an experiment confirming this finding with the currently used fee values.
We simulate a workload of demand in the $L$-to-$R$ direction: 60000 $L$-to-$R$ and 15000 $R$-to-$L$ transactions under a high (10 tx/minute $L$-to-$R$, 2.5 tx/minute $R$-to-$L$) and a low (1 tx/minute $L$-to-$R$, 0.25 tx/minute $R$-to-$L$) intensity.
The node's total fortune over time for high and low intensity are shown in Figs. \ref{fig:results_202_total_fortune_over_time} and \ref{fig:results_204_total_fortune_over_time} respectively.
We see that regardless of the (non-\textit{None}) rebalancing policy, the node's fortune decreases over time, because rebalancing fees surpass any relay profits, which are small because of the small $f_{\mathrm{prop}}$ compared to $F$.
In this regime, the node is better off not rebalancing at all.
Still, our RebEL policy manages to learn this fact and after some point exhibits the desired behavior and stops rebalancing as well.
Autoloop and Loopmax keep trying to rebalance and end up exhausting their entire on-chain balance, so the total fortune under them gets stuck after some point.

Taking a higher level view, we also conduct multiple experiments with the same demand as before but now while varying $f_{\mathrm{prop}}$.
The results of the total final fortune of each experiment (run for the same total time and averaged over 10 runs; error bars show the maximum and minimum values) are shown in Fig. \ref{fig:results_212_final_fortune_wrt_proportional_fee} under high demand and in Fig. \ref{fig:results_214_final_fortune_wrt_proportional_fee} under low demand.
We see that no rebalancing policy is profitable (i.e. better than \textit{None}) as long as $f_{\mathrm{prop}} < 0.5\% = F$, which confirms our back-of-the-envelope calculation.
For higher values of $f_{\mathrm{prop}}$, the node is able to make a profit.
Although RebEL performs better for $f_{\mathrm{prop}}=1\%$ for reasons discussed in Sec. \ref{sec:experiments-wrt-demand}, Autoloop and Loopmax sometimes perform better for even higher (and thus even farther from the current) fees, because the RebEL policy used in this experiment is the one we tuned to operate best for the experiments of the next section that use $f_{\mathrm{prop}}=1\%$.
In principle though, with different tuning, RebEL could outperform the other policies for higher values of $f_{\mathrm{prop}}$ as well.

\subsubsection{The role of the demand}
\label{sec:experiments-wrt-demand}

\begin{figure}[ht]
    \centering
    \subfigure[Total fortune over time]{
        \includegraphics[width=0.31\textwidth]{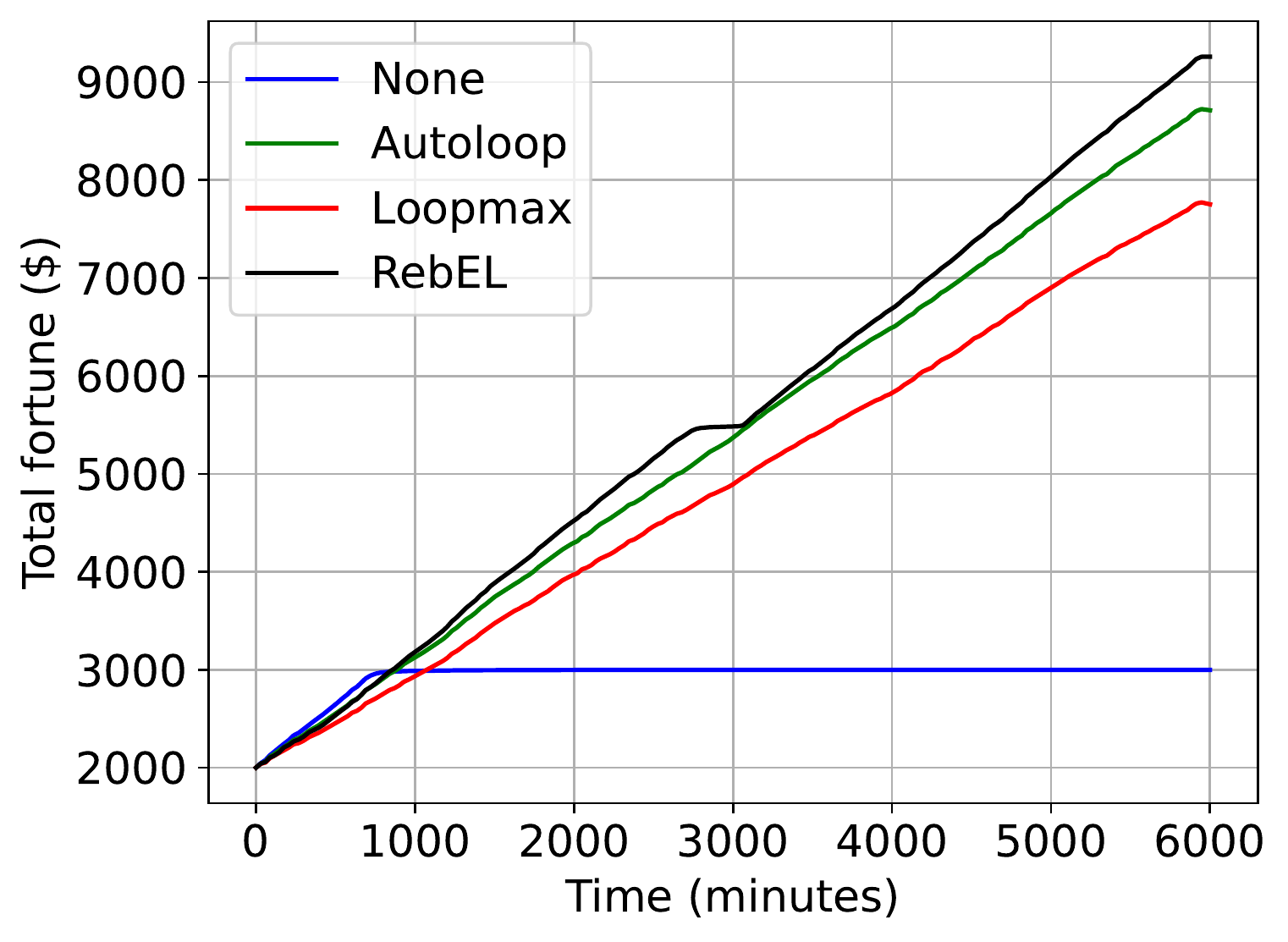}
        \label{fig:results_102_total_fortune_over_time}
    }
    \subfigure[Transaction fee losses over time]{
        \includegraphics[width=0.31\textwidth]{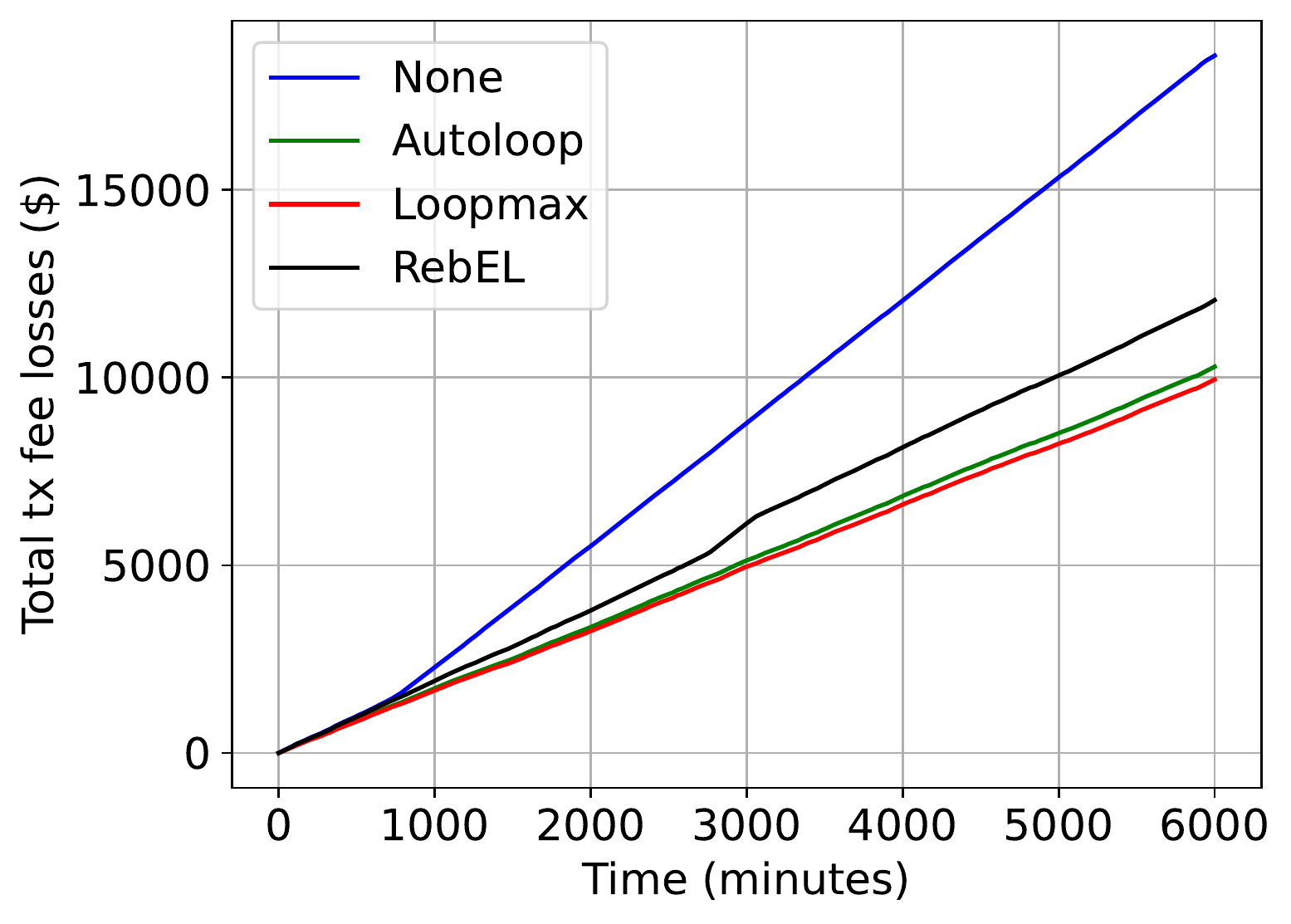}
        \label{fig:results_102_tx_fee_losses_over_time}
    }
    \subfigure[Rebalancing fees over time]{
        \includegraphics[width=0.31\textwidth]{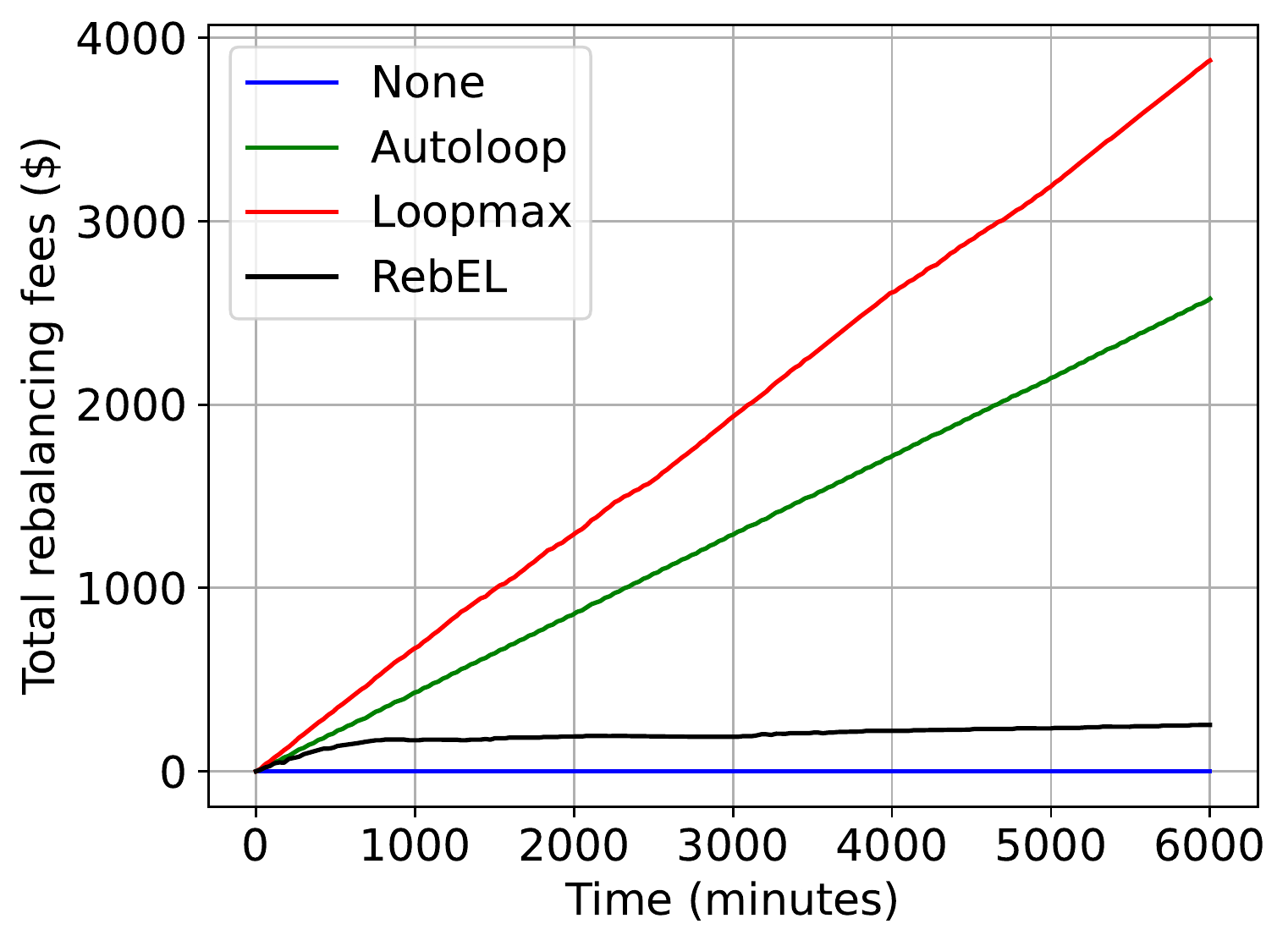}
        \label{fig:results_102_reb_fees_over_time}
    }
    \caption{Total fortune, transaction fee losses and rebalancing fees over time under high demand skewed in the $L$-to-$R$ direction}
    \label{fig:demand-skewed-high}
\end{figure}

\begin{figure}[ht]
    \centering
    \subfigure[Total fortune over time]{
        \includegraphics[width=0.31\textwidth]{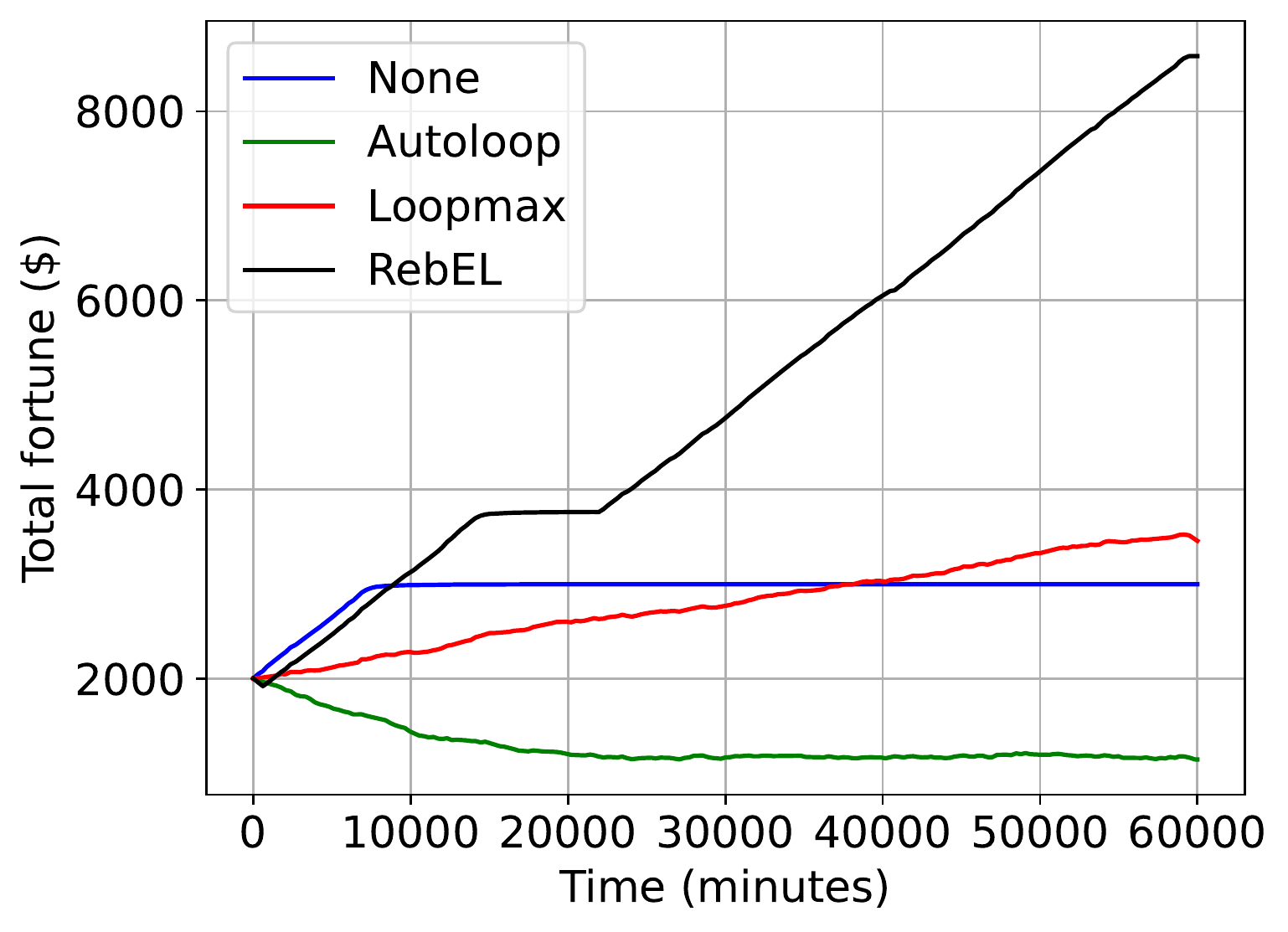}
        \label{fig:results_104_total_fortune_over_time}
    }
    \subfigure[Transaction fee losses over time]{
        \includegraphics[width=0.31\textwidth]{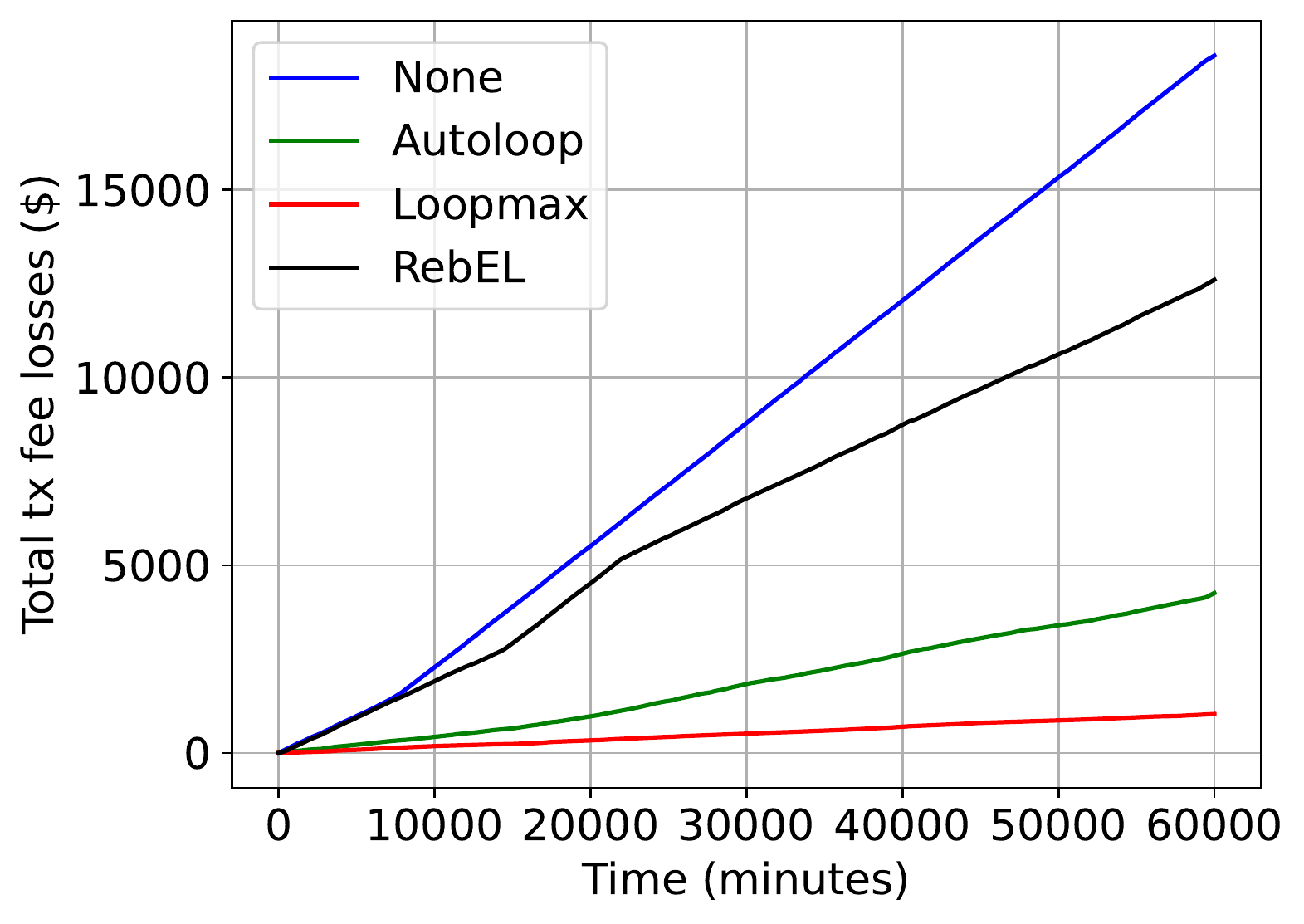}
        \label{fig:results_104_tx_fee_losses_over_time}
    }
    \subfigure[Rebalancing fees over time]{
        \includegraphics[width=0.31\textwidth]{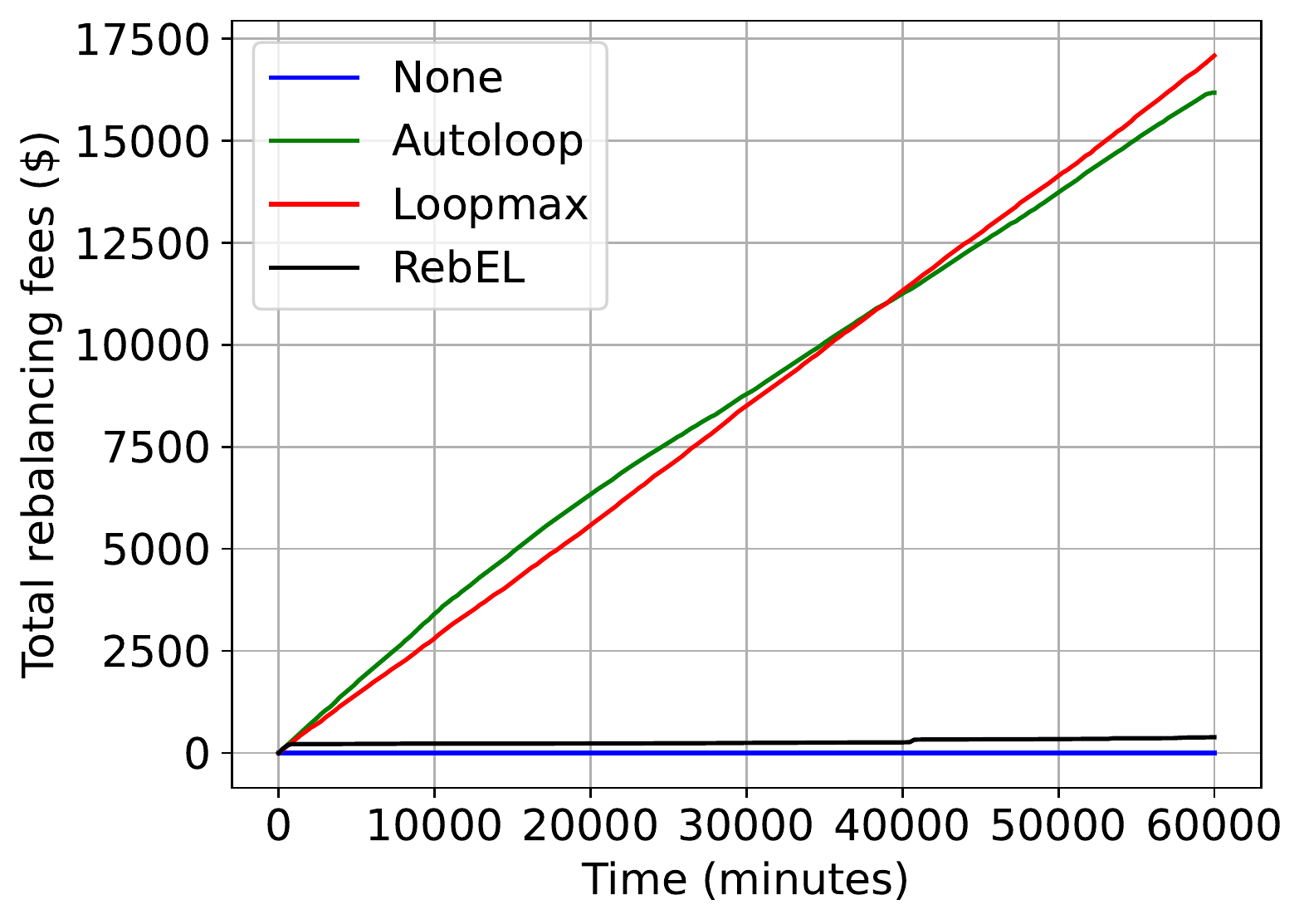}
        \label{fig:results_104_reb_fees_over_time}
    }
    \caption{Total fortune, transaction fee losses and rebalancing fees over time under low demand skewed in the $L$-to-$R$ direction}
    \label{fig:demand-skewed-low}
\end{figure}

We now stay in the fee regime of possible profitability, i.e. by keeping $f_{\mathrm{prop}}=1\%$, and try to understand the role of the demand (and indirectly of the depletion frequency) on the performance of the different policies.
The results for the same high and low workload of skewed demand in the $L$-to-$R$ direction as before are shown in Figs. \ref{fig:demand-skewed-high} and \ref{fig:demand-skewed-low}.


RebEL outperforms all other policies under both demand regimes (Figs. \ref{fig:results_102_total_fortune_over_time}, \ref{fig:results_104_total_fortune_over_time}), as it manages to strike a balance in terms of frequency and amount of rebalancing and transaction fee profits.
This happens in a few 10-minute iterations under high demand (corresponding to a few hours in real time), because balance changes are more pronounced in this case and help RebEL learn faster, while it takes about 1200 iterations under low demand, translating in 8.3 days of training.
Both these training times are reasonable for a relay node investing its capital to make a profit.
We see that under both regimes the system without rebalancing (\textit{None} policy) at some point reaches a state where almost all the balances are accumulated locally and no transactions can be processed anymore (hence the flattening in the \textit{None} curve).
Under high demand, Autoloop and Loopmax rebalance a lot (Fig. \ref{fig:results_102_reb_fees_over_time}) in order to minimize transaction fee losses (Fig. \ref{fig:results_102_tx_fee_losses_over_time}), while RebEL sacrifices some transactions to achieve higher total fortune.
Under low demand, RebEL rebalances only when necessary (Fig. \ref{fig:results_104_reb_fees_over_time}), even if this means sacrificing many more transactions (Fig. \ref{fig:results_104_tx_fee_losses_over_time}), simply because rebalancing is not worth it at that low demand regime, in the sense that the potential profits during the 10-minute rebalancing check times are too low to justify the frequent rebalancing operations that the other policies apply.
Loopmax eventually achieves a profit (although much lower than RebEL) because it tends to rebalance with higher amounts.
On the contrary, Autoloop rebalances with small amounts, thus incurring significant costs from constant miner fees and eventually even making a loss compared to the initial node's fortune (Fig. \ref{fig:results_104_total_fortune_over_time}).
Under high demand, there is a point around time 2700 where RebEL stalls for a bit, and the same happens under low demand between times 14000-22000.
Upon more detailed inspection, this happens because all balances temporarily accumulate on the local sides of the channels.
RebEL takes some steps to again bring the channels to some balance (either actively by making a swap or passively by letting transactions flow) and subsequently completely recovers.

\begin{wrapfigure}{R}{0.32\textwidth}
    \centering
    \subfigure[High demand]{
        \includegraphics[width=0.3\textwidth]{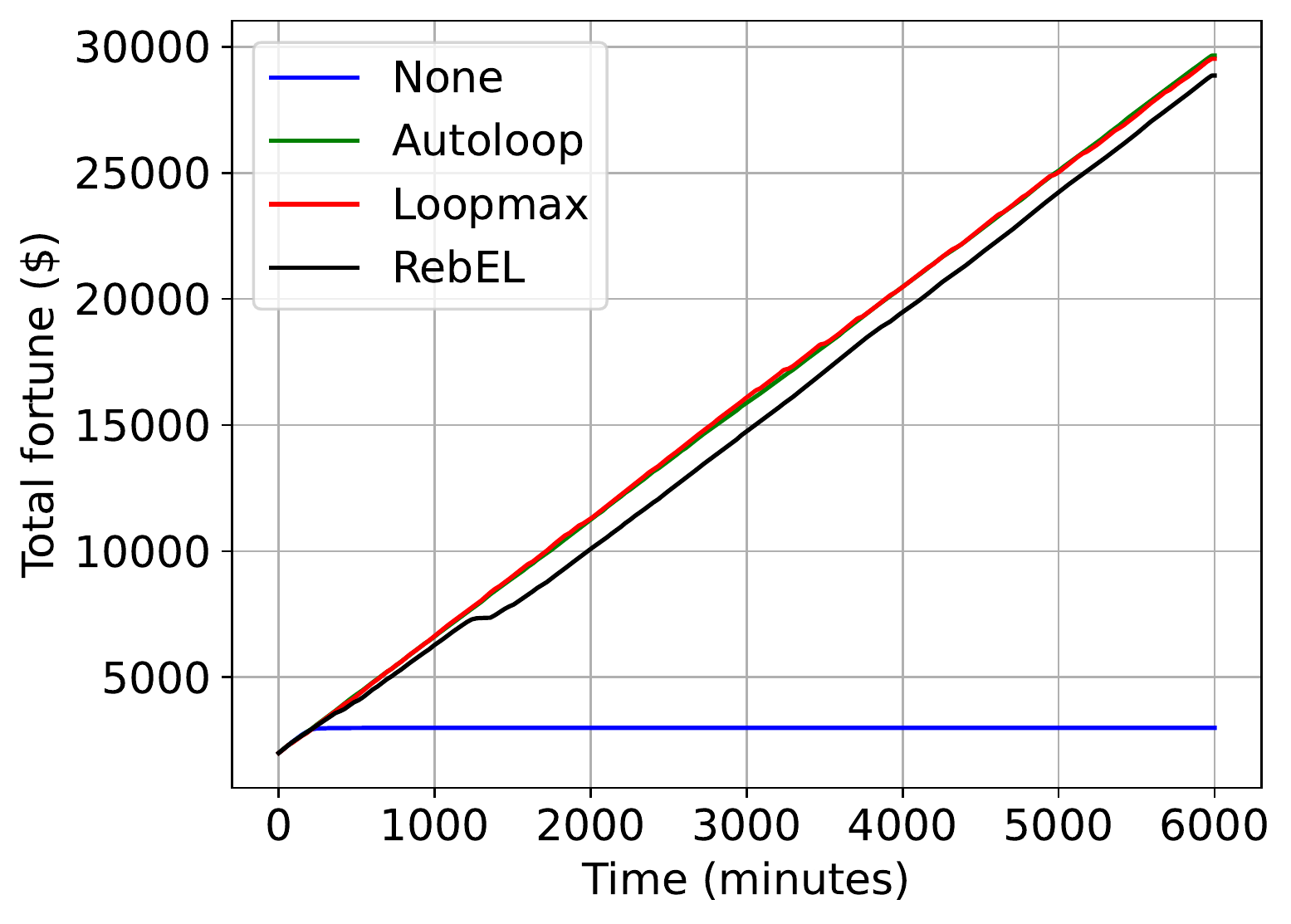}
        \label{fig:results_106}
    } \\
    \subfigure[Low demand]{
        \includegraphics[width=0.3\textwidth]{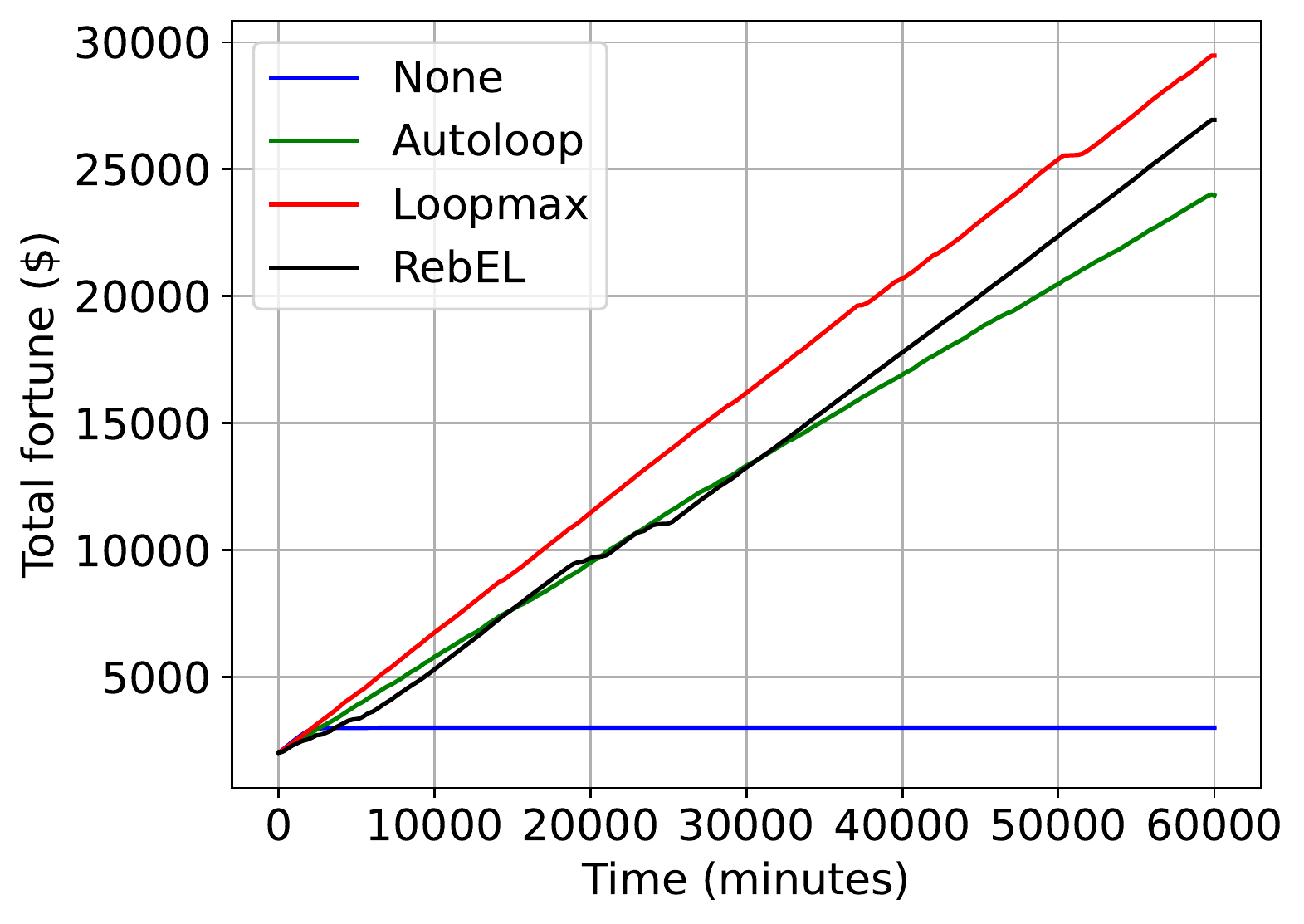}
        \label{fig:results_108}
    }
    \caption{Total fortune over time under equal demand intensity from both sides}
    \label{fig:demand-high-vs-low-even}
\end{wrapfigure}

We now explore the special case of equal demands, by applying 60000 transactions arriving on each side in high (10 tx/minute) and low (1 tx/minute) intensity.
Tuning some hyperparameters and making the penalty for failed swaps non-zero as shown in Appendix \ref{app:hyperparameters-and-rewards} gave better results for even demand specifically, so we use this configuration for the results of Fig. \ref{fig:demand-high-vs-low-even}.
We observe that all policies (except \textit{None}) achieve higher total fortunes than before.
This happens because the almost even traffic automatically rebalances the channel to some extent and therefore more fees can be collected in both directions and for larger amounts of time before the channels get stuck.
RebEL is not as good for even traffic, because the net demand constantly oscillates around zero and this does not allow the agent to learn a good policy.
It still manages though to surpass Autoloop pretty quickly under low demand, while if we run the simulation for longer times (not shown in the figure), we see that after time 78000 RebEL surpasses Loopmax as well. 
This translates to about 54 days of operation, which is a big time interval in practice, but is justified by the fact that the traffic is low and therefore more time is needed in order for the node to make a profit.
However, even demand from both sides is a special case that is not likely to occur in practice, as usually the traffic follows some patterns, e.g. from clients to merchants.
So the skewed demand scenario, where RebEL is superior, is also the most natural.

\subsubsection{The role of initial conditions}
\label{sec:experiments-wrt-initial-conditions}

\begin{figure}[ht]
    \centering
    \subfigure[Total fortune over time when $C_L = 1000$, $C_R = 500$]{
        \includegraphics[width=0.31\textwidth]{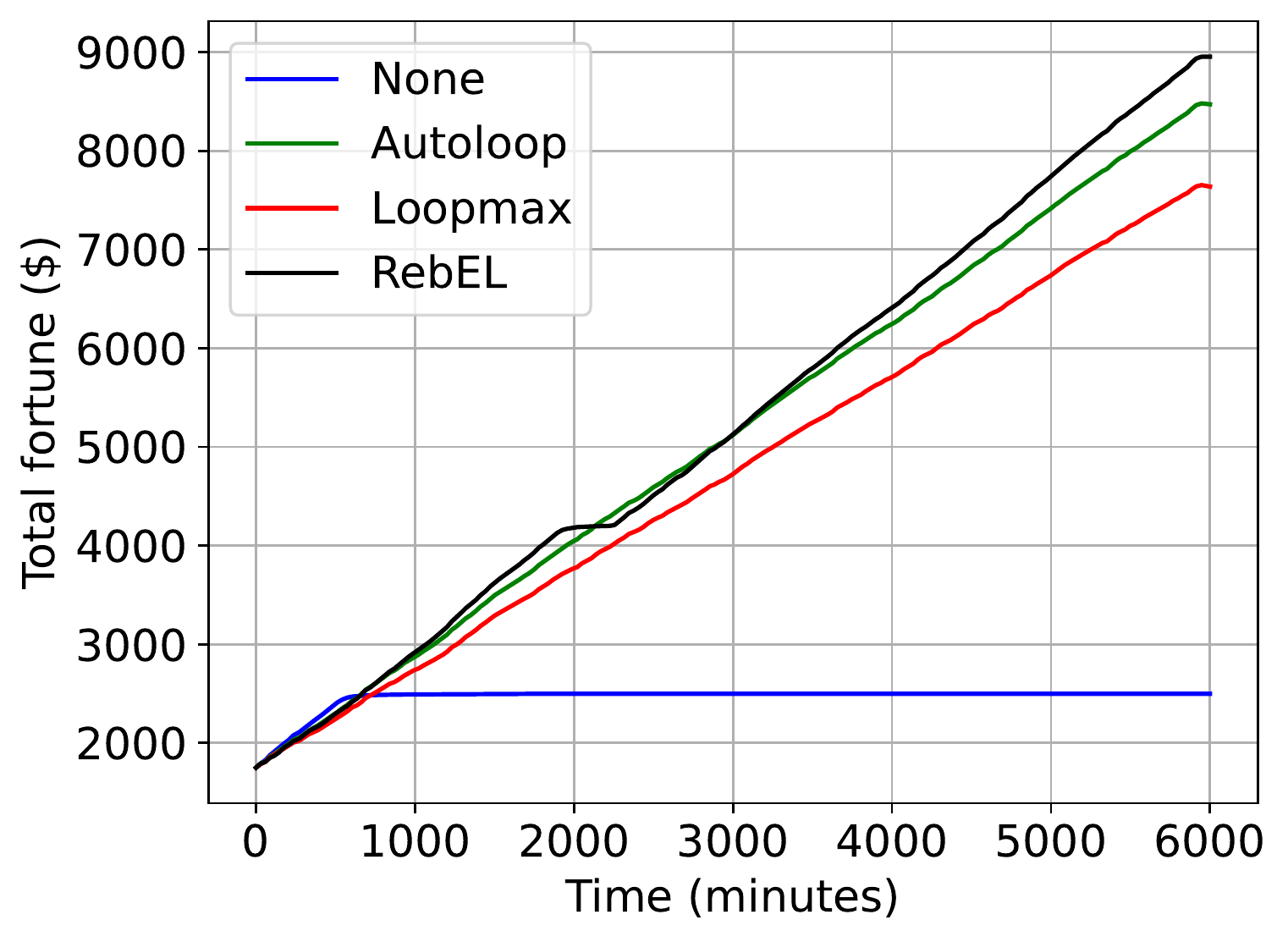}
        \label{fig:results_302a_total_fortune_over_time}
    }
    \subfigure[Total fortune over time when $C_L = 500$, $C_R = 1000$]{
        \includegraphics[width=0.31\textwidth]{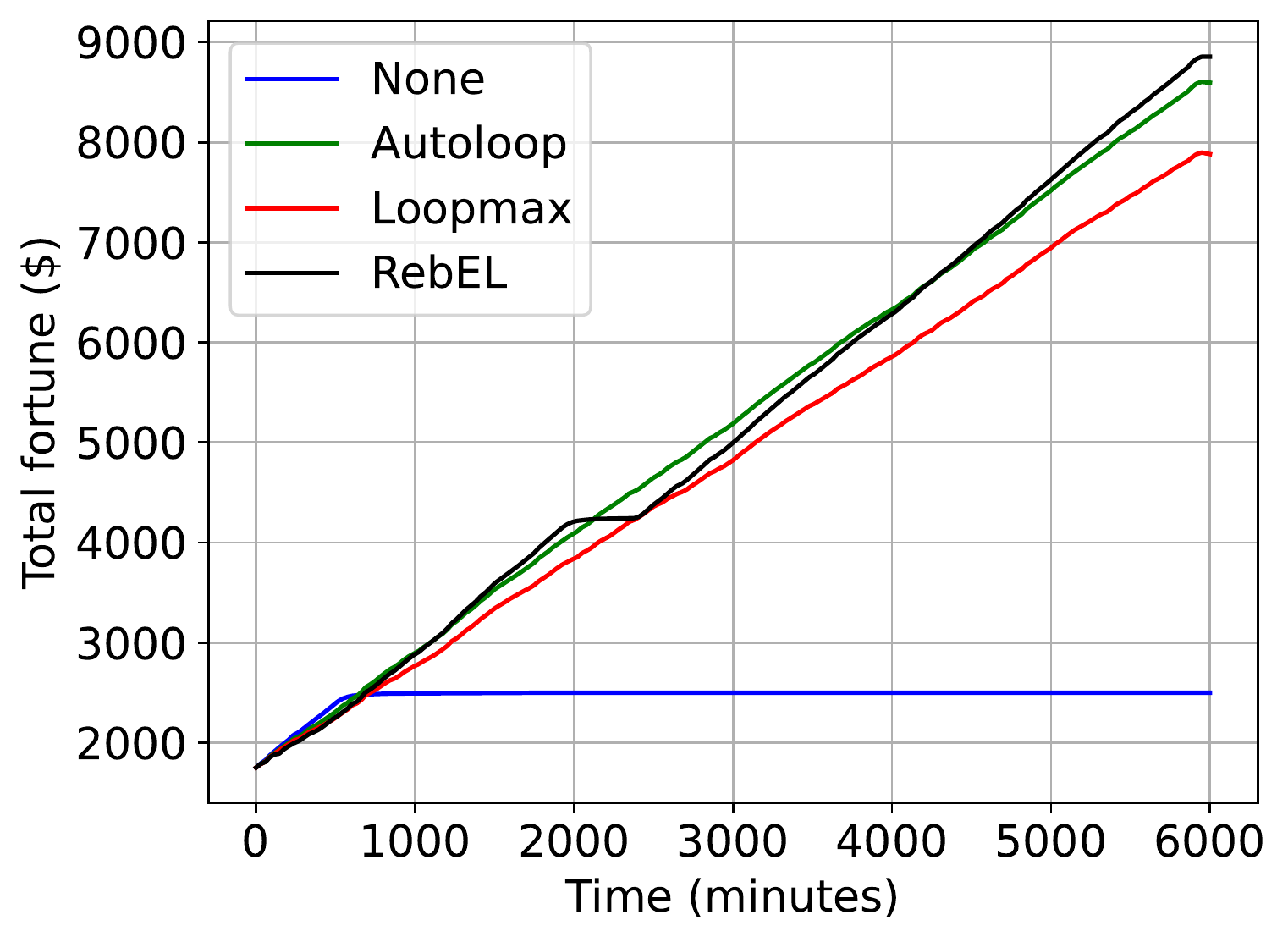}
        \label{fig:results_302b_total_fortune_over_time}
    }
    \subfigure[Total fortune over time when initial balances are only local]{
        \includegraphics[width=0.31\textwidth]{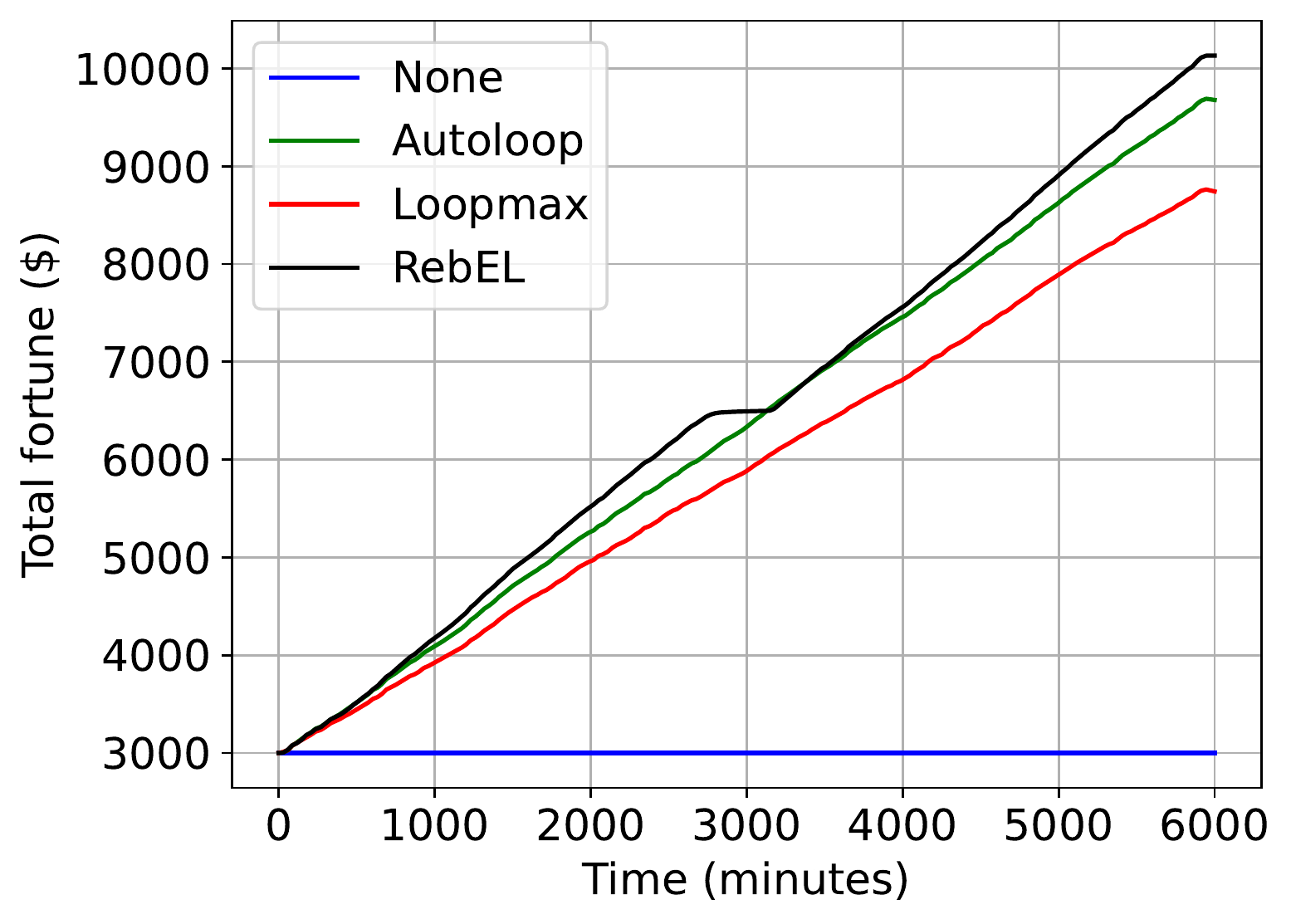}
        \label{fig:results_302c_total_fortune_over_time}
    }
    \caption{Total fortune, transaction fee losses and rebalancing fees over time under high demand skewed in the $L$-to-$R$ direction for different initial conditions}
    \label{fig:initial-conditions-high-demand}
\end{figure}

\begin{figure}[ht]
    \centering
    \subfigure[Total fortune over time when $C_L = 1000$, $C_R = 500$]{
        \includegraphics[width=0.31\textwidth]{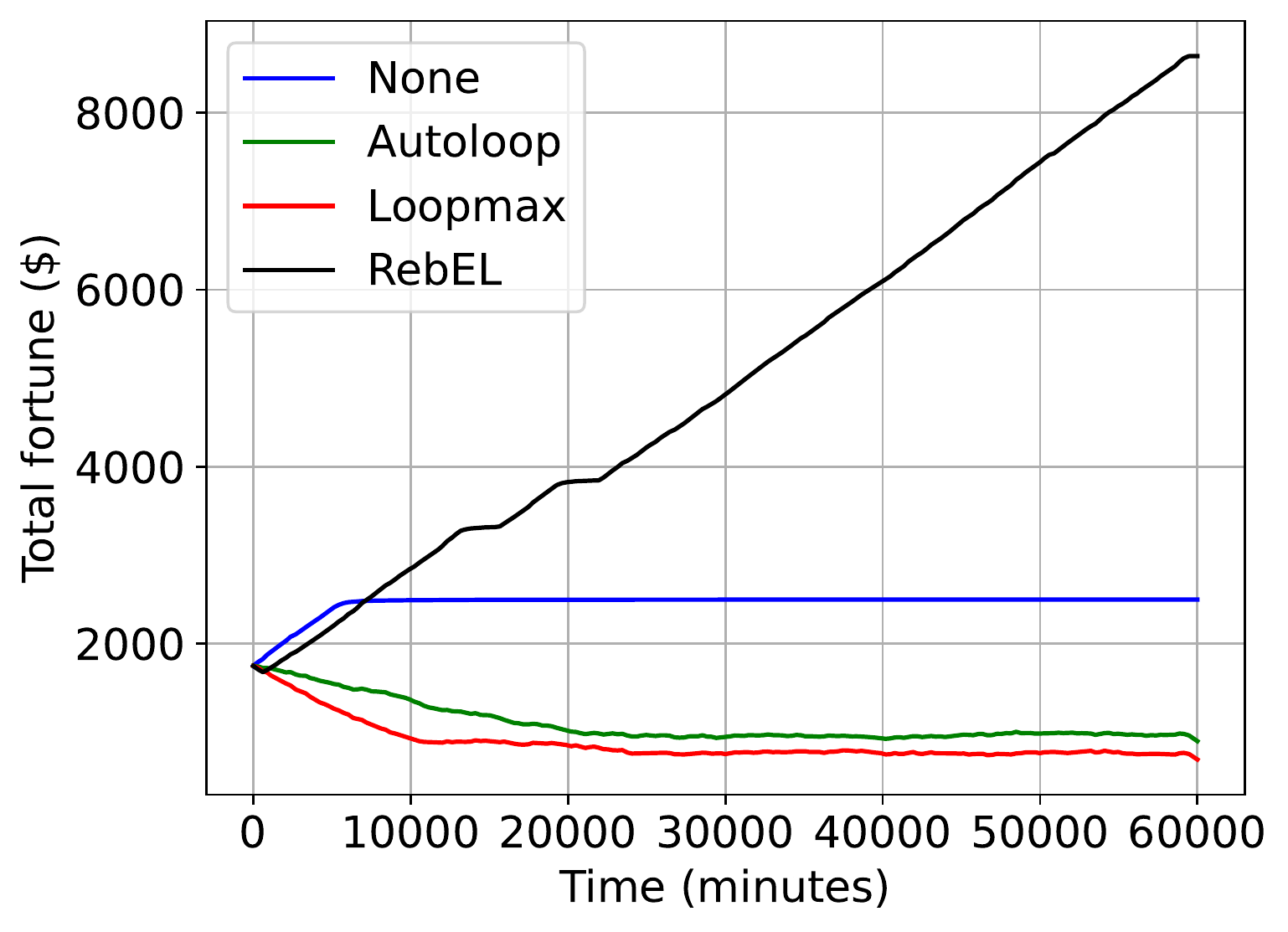}
        \label{fig:results_304a_total_fortune_over_time}
    }
    \subfigure[Total fortune over time when $C_L = 500$, $C_R = 1000$]{
        \includegraphics[width=0.31\textwidth]{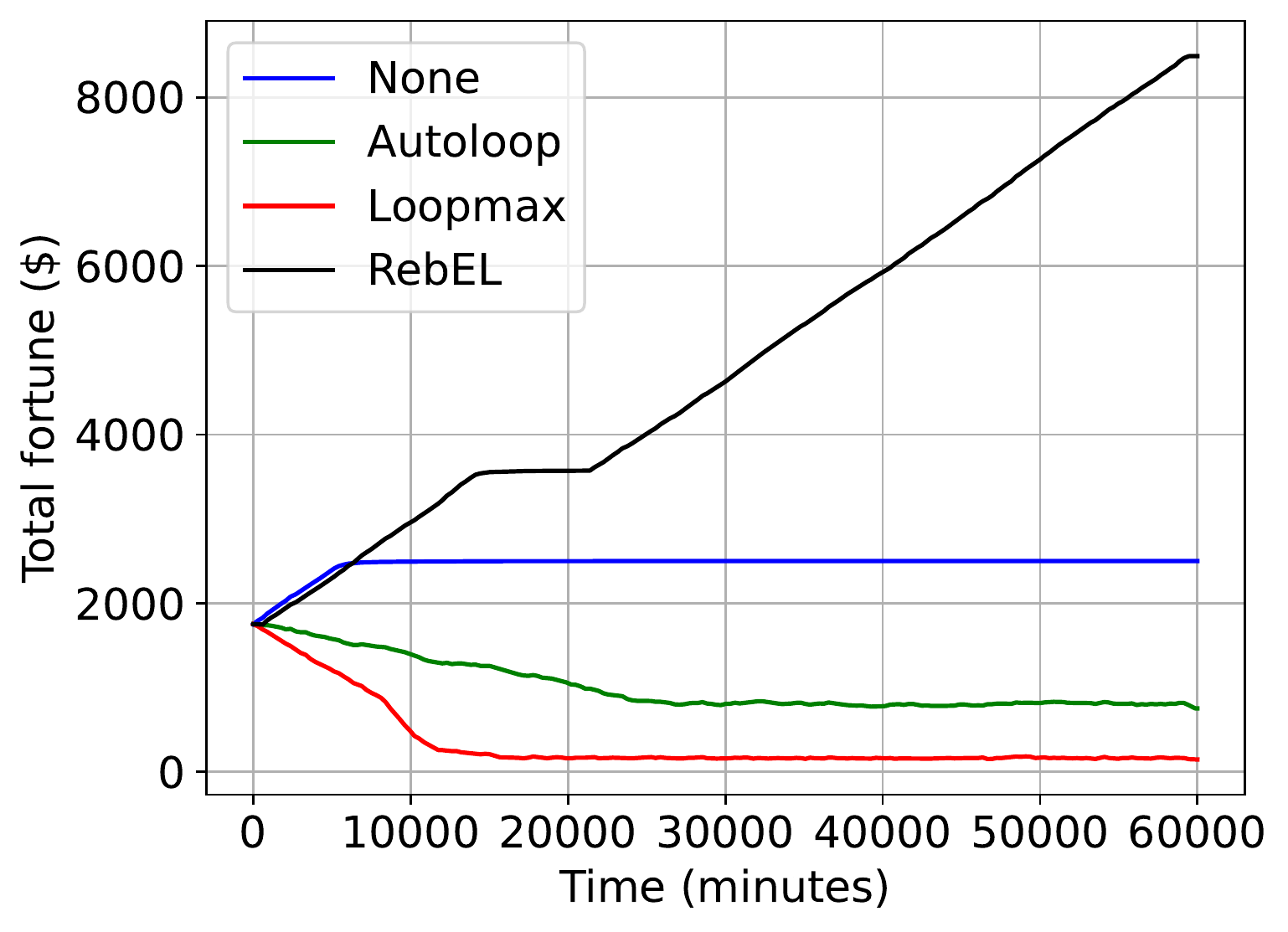}
        \label{fig:results_304b_total_fortune_over_time}
    }
    \subfigure[Total fortune over time when initial balances are only local]{
        \includegraphics[width=0.31\textwidth]{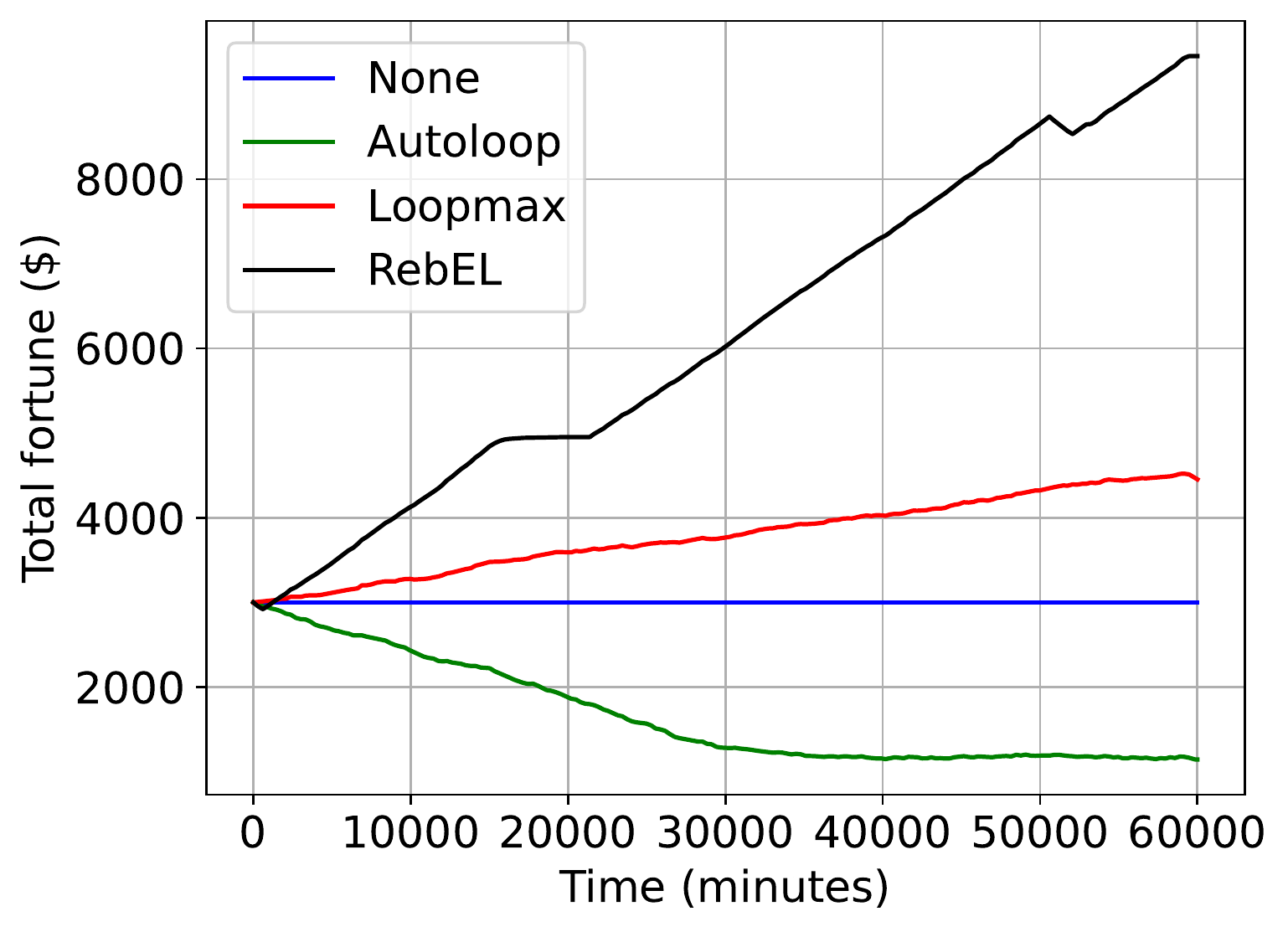}
        \label{fig:results_304c_total_fortune_over_time}
    }
    \caption{Total fortune, transaction fee losses and rebalancing fees over time under low demand skewed in the $L$-to-$R$ direction for different initial conditions}
    \label{fig:initial-conditions-low-demand}
\end{figure}

We now examine how the initial conditions (capacities, balances) affect the performance.
We evaluate all rebalancing policies for the skewed demand in the $L$-to-$R$ direction scenario as before, but this time for channels of uneven capacities or initial balances.
The results for high and low demand are shown in Figs. \ref{fig:results_302a_total_fortune_over_time} and \ref{fig:results_304a_total_fortune_over_time} respectively for $C_L = 1000$, $C_R = 500$ and the initial balances evenly distributed, in Figs. \ref{fig:results_302b_total_fortune_over_time} and \ref{fig:results_304b_total_fortune_over_time} respectively for $C_L = 500$, $C_R = 1000$ and the initial balances evenly distributed, and in Figs. \ref{fig:results_302c_total_fortune_over_time} and \ref{fig:results_304c_total_fortune_over_time} respectively for $C_L = C_R = 1000$ but $b_{NL} = b_{NR} = 1000$ (and so $b_{LN} = b_{RN} = 0$).
We see that RebEL performs well in all these cases as well.
Depending on the exact arriving transactions, the little plateaus of RebEL happen at different points in time for the same reason as before, but in the end the learning algorithm recovers.

\newpage
\section{Discussion and future work}
\label{sec:discussion}

We now make some remarks on the design and the practical applicability of our DRL-based policy and discuss future extensions of our work.

\par{\textit{Design choices}}:
The objectives of Section \ref{sec:writing-as-an-MDP} were defined as long-term expected average ones in order to match what a relay node would intuitively want to optimize, while the SAC algorithm works for long-term discounted objectives with a discount factor (usually set very close to 1), and including a maximum entropy term to enhance exploration\footnote{The exact formula for the SAC objective can be found in Appendix A of \cite{SAC-paper-2}.}.
We expect this difference to not be significant, and indeed the results show that the SAC-based policy performs well in practice.
Furthermore, in Sec. \ref{sec:evaluation} we presented results for specific parameters and rewards for the RL algorithm.
Further tuning specific to the demand regime might lead to even higher returns for the RebEL policy.
Additionally, improving the estimates of future balances by having the agent perform a ``mini-simulation'' of the transactions arriving in the following time interval based on past statistics could help the policy produce more informed decisions.
Techniques from Model Predictive Control could also be applied \cite{Rosolia2018}.

Theoretically, a class of policies that could result in even higher fortune than the class \eqref{eqn:admissible-policies} would be one that would allow rebalancing to happen at any point in continuous time instead of periodically. 
Optimization in such a model however would be extremely difficult, as an action taken now would affect the state both now and in the future (when rebalancing completes).
Considering that practical policies like Autoloop applied today only check for rebalancing periodically, we follow the same path for the sake of tractability.

\par{\textit{Practical applicability}}:
An actual PCN node could use our simulator with samples from its past demand, and try to tune the RL parameters and the reward to get better performance than the heuristic policies we defined or the one it is currently using; then, it would apply the policy learned in the simulator environment to the real node.
Alternatively, a node may not use a simulator at all and directly learn a pre-parameterized policy on the fly from the empirical transaction data.
In either case, the node can do occasional retraining with updated data to account for time-variance in the distribution of the arriving demand.

\par{\textit{Future directions}}:
Our two-channel DRL solution was a proof of concept that DRL can indeed be applied for liquidity management in PCNs.
Armed with this knowledge, in future research we intend to study the more general case of a node being the center of a star graph of channels and trying to make a profit while rebalancing all of them appropriately.
Another extension would be to allow the node to batch rebalancing operations into one on-chain transaction to save on on-chain fees.
Moreover, in our work we considered the neighboring nodes $L$ and $R$ to be passive.
Future work can investigate a game-theoretic framework where all PCN nodes are rational and compete against each other towards making a profit.
Finally, it would also be interesting to compare the performance of different rebalancing methods, depending on the demand and channel conditions.

\section{Related work}
\label{sec:related-work}

\par{\textit{Rebalancing methods}}:
Rebalancing via payments from a node to itself via a circular path of channels has been studied by \cite{Khalil2017, Pickhardt2020, Avarikioti2021, Xu2021, Hong2022, Bastankhah2022}.
Some of them take relay fees into account as we did, and some do not.
\cite{Awathare2021} in particular performs circular rebalancing coupled with a rerouting scheme based on a metric that accounts for the average demand in a simple way (we did so too in defining the different balance estimates).
\cite{DiStasi2018, vanEngelshoven2021} describe fee strategies that incentivize the balanced use of payment channels.
\cite{Bai2022} uses a game-theoretic lens to study the extent to which nodes can pay lower transaction fees by waiting patiently and reordering transactions instead of pursuing maximum efficiency.
Perhaps the only work on submarine swaps, \cite{Galoy2022}, shows that there is a possibility of liquidity arbitrage of Lightning liquidity providers by users, which in turn determines a market rate for acquiring liquidity, and then develops fee structures for properly pricing liquidity without overcharging regular users.
In \cite{Ersoy2020}, a more holistic view is attempted regarding an optimization decision a blockchain node with an initial budget has to make: how to maximize the average gain per incoming transaction from a \textit{known} distribution by choosing which channels to open, with what capacities and with what fees.
However, the model ignores the channel opening costs by assuming it is possible to extend the channel's lifetime arbitrarily, without though detailing how this would be done (e.g. via rebalancing).
A recent development similar to submarine swaps is PeerSwap \cite{peerswap-1, peerswap-2}: instead of buying funds from an LSP, a node can exchange funds on-/off-chain with its channel neighbor directly.
\textit{Splicing} is another mechanism that replaces a channel with a new one with a different capacity while allowing transactions to flow in the meantime \cite{splicing}.

\par{\textit{Techniques}}: Stochastic modeling and optimization in the blockchain space has been used both in layer-1 \cite{Dembo2020, Gazi2020, Papadis2018, Misic2020} for performance characterization, and in layer-2 for routing \cite{Varma2021} and scheduling \cite{Papadis2021} of payments.
Deep Reinforcement Learning has been broadly applied to approximately solve challenging optimization problems from various areas and to build systems that learn to manage resources directly from experience.
For example, \cite{alizadeh2016} applies DRL to the resource allocation problem of packing tasks under multiple resource demands, while \cite{Bar-Zur2022} describes a DRL framework for solving a complex MDP underlying the incentives around selfish mining attacks in Bitcoin-like blockchains.
Our profitable rebalancing problem resembles problems appearing in stochastic inventory control, without or with a positive lead time for replenishment.
The so-called $(s,S)$ threshold policies (if inventory level $x<s$, order $S-x$; if $x>s$, do not order) can be proved to be optimal in certain settings \cite{Scarf1959, Zheng1991, Sethi1997}.
Autoloop resembles these policies; however, our problem presents additional complexities due to the fact that there are more than one channels, with the balances of each affecting transaction processing in the other, leading us to a DRL-based approach.
(Deep) RL has been applied extensively to inventory management problems as well \cite{VanRoy1997, DRL-inventory-1}, although usually extensive tuning is necessary \cite{DRL-inventory-2}.

\section{Conclusion}
\label{sec:conclusion}

In this paper, we studied the problem of relay node profit maximization using submarine swaps, and demonstrated the feasibility of applying state-of-the-art DRL techniques for solving it, with our experiments showing that a SAC-based policy can outperform heuristic policies in most cases.
We hope that this research will inspire further interest in designing capital management strategies in the complex world of PCNs based on learning from experience as an alternative to currently applied heuristics, and will be a step towards guaranteeing the profitability of the relay nodes and, consequently, the viability and scalability of the PCNs they sustain.

\printbibliography[
    title={References},
]

\appendix

\newpage

\section{Example of channel depletion under symmetric demand}
\label{app:symmetric-depletion}

\begin{figure}[h]
    \centering
    \includegraphics[width=0.4\textwidth]{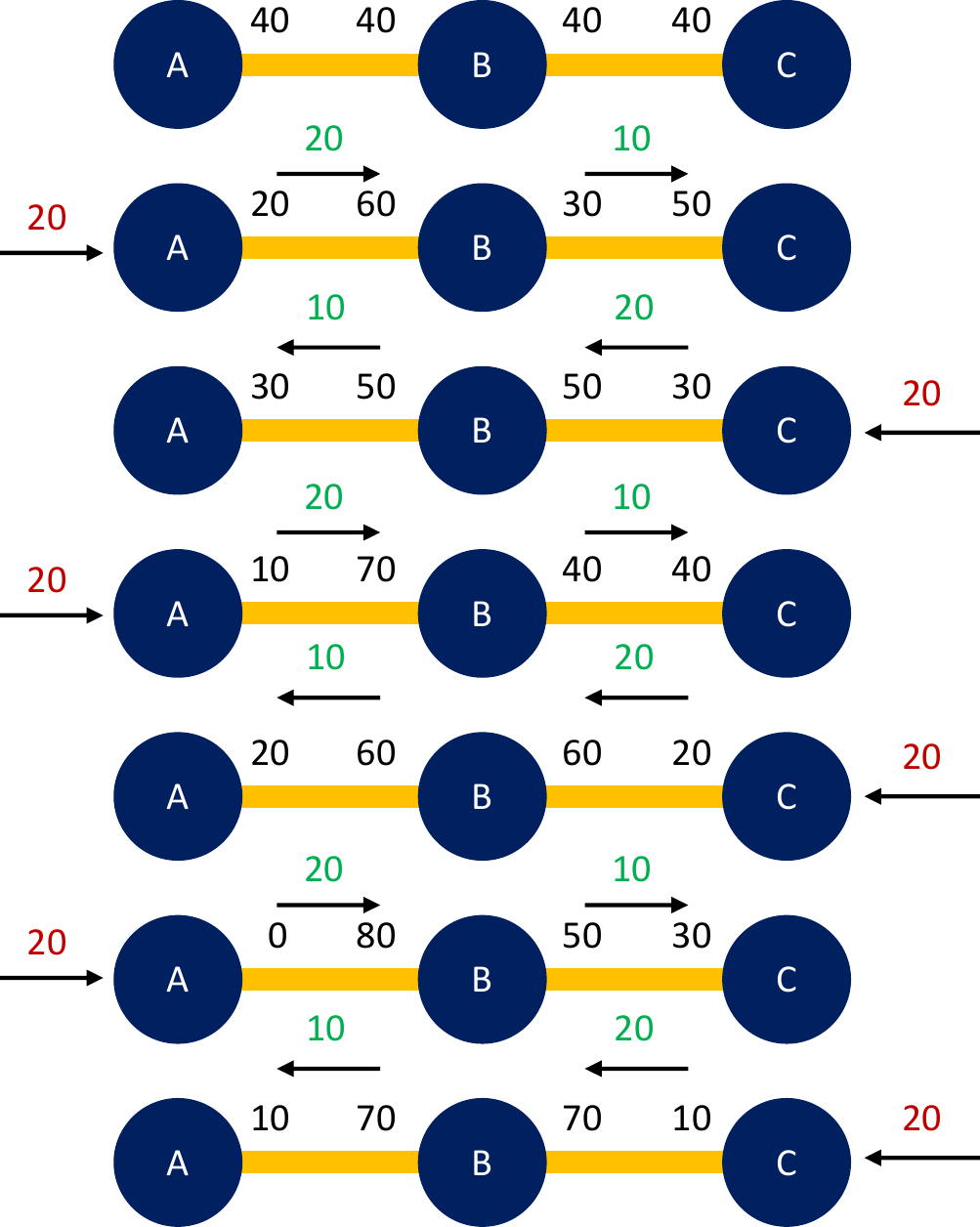}
    \caption{An example of a PCN getting stuck even though the demand is symmetric. Demand is shown in red, forwarded amounts after a 50\% fee withholding are shown in green, and channel balances are shown in black.}
    \label{fig:symmetric_demand_depletion_example}
\end{figure}

Symmetric demand on two endpoints of a multihop path can cause imbalance due to fees withheld by intermediate nodes.
Fig. \ref{fig:symmetric_demand_depletion_example} shows the evolution over time of a subnetwork of three channels with symmetric demand of amount 20 arriving alternately from either side of the path.
When each transaction is relayed by node $B$, a 50\% fee is withheld and the remaining amount of 10 is forwarded to the next channel in the path.
We see that even though the end-to-end path demand is symmetric, after a few steps the channels get unbalanced and stop being able to process any more transactions\footnote{
The 50\% fee is not realistic and is only used for the purposes of this example.
With the real much lower fees the channels will similarly get stuck after a larger number of steps.
}.

\newpage

\section{Hyperparameters and rewards}
\label{app:hyperparameters-and-rewards}

\begin{table}[h!]
\centering
\caption{SAC hyperparameters used for the different experiments of Sec. \ref{sec:evaluation}}
\label{table:sac-hyperparameters}
\makebox[\textwidth]{\begin{NiceTabular}{cwc{0.28\textwidth}wc{0.28\textwidth}}[hvlines,corners=NW, cell-space-limits=2pt] 
    \textbf{SAC hyperparameter}     
        & \Block[]{}{\textbf{Parameter value}\\\textbf{for skewed demand}\\\textbf{experiments}}
        & \Block[]{}{\textbf{Parameter value}\\\textbf{for even demand}\\ \textbf{experiments}} \\    
    policy                  & \Block[c]{1-2}{Gaussian}  \\
    optimizer               & \Block[c]{1-2}{Adam} \\ 
    learning rate           & 0.0003    & 0.006 \\ 
    discount                &  \Block[c]{1-2}{0.99} \\
    replay buffer size      &  \Block[c]{1-2}{$10^5$} \\
    \Block{}{number of hidden layers\\(all neural networks)} & \Block[c]{1-2}{2}  \\
    number of hidden units per layer                    & \Block[c]{1-2}{256}  \\
    number of samples per minibatch                     & \Block[c]{1-2}{10} \\
    temperature             & 0.05                      & 0.005  \\
    nonlinearity                        & \Block[c]{1-2}{ReLU}   \\
    target smoothing coefficient        & \Block[c]{1-2}{0.005}  \\
    target update interval              & \Block[c]{1-2}{1}  \\
    gradient steps                      & \Block[c]{1-2}{1}  \\
    automatic entropy tuning            & False         & True  \\
    initial random steps                & \Block[c]{1-2}{10}  \\
    \end{NiceTabular}}
\end{table}

\begin{table}[h!]
\centering
\caption{Parameters used in RebEL's representation or processing of the states, actions, and rewards}
\label{table:RebEL-parameters}
\begin{NiceTabular}{cwc{0.28\textwidth}wc{0.28\textwidth}}[hvlines,corners=NW, cell-space-limits=2pt] 
    \textbf{RebEL parameter}     
        & \Block[]{}{\textbf{Parameter value}\\\textbf{for skewed demand}\\\textbf{experiments}}
        & \Block[]{}{\textbf{Parameter value}\\\textbf{for even demand}\\ \textbf{experiments}} \\  
    \Block{}{on-chain amount\\normalization constant}  & \Block{1-2}{60}   \\ 
    minimum swap threshold $\rho_0$             & \Block{1-2}{0.2}  \\ 
    penalty per swap failure                    & 0         & 10  \\
    \end{NiceTabular}
\end{table}

\end{document}